\documentclass[preprint,3p,12pt]{elsarticle}
\usepackage{srcltx}
\usepackage{times,amsbsy,amsmath,amsthm,amssymb,amscd,amsfonts,mathrsfs}
\usepackage{graphicx,epstopdf}
\usepackage[hang]{subfigure}
\usepackage{threeparttable}
\usepackage{multirow}
\usepackage{dcolumn}
\usepackage{booktabs}
\usepackage{tikz}
\usetikzlibrary{arrows, snakes, backgrounds}
\usepackage{float}
\usepackage{color}
\usepackage{macros}
\usepackage{algorithm}
\usepackage{algorithmic}

\newcommand{\pd}[2]{\dfrac{\partial #1}{\partial #2}}
\newcommand{\dd}{\,{\rm d}}

\def\ie{\mathrm{e}}

\usepackage{lineno,hyperref}
\modulolinenumbers[5]

\journal{Journal of \LaTeX\ Templates}

\begin{document}

\begin{frontmatter}

\title{ Unified Gas-kinetic Wave-Particle Methods III: Multiscale
Photon Transport}

\author[ad1,ad3]{Weiming Li}
\ead{liweiming@csrc.ac.cn}
\author[ad3]{Chang Liu}
\ead{cliuaa@connect.ust.hk}
\author[ad2]{Yajun Zhu}
\ead{zhuyajun@mail.nwpu.edu.cn}
\author[ad1]{Jiwei Zhang}
\ead{jwzhang@csrc.ac.cn}
\author[ad3,ad4]{Kun Xu\corref{cor1}}
\ead{makxu@ust.hk}
\address[ad1]{Applied and Computational Mathematics
Division, Beijing Computational Science Research Center, Beijing 100193, China}
\address[ad2]{National Key Laboratory of Science and Technology on Aerodynamic Design and Research, Northwestern Polytechnical University, Xi'an, Shaanxi 710072, China}
\address[ad3]{Department of Mathematics, Hong Kong University of
Science and Technology, Clear Water Bay, Kowloon, Hong Kong, China}
\address[ad4]{HKUST Shenzhen Research Institute, Shenzhen 518057, China}
\cortext[cor1]{Corresponding author}

\begin{abstract}
In this paper, we extend the unified gas-kinetic wave-particle (UGKWP)
method to the multiscale photon transport. In this method, the photon
free streaming and scattering processes are treated in an un-splitting
way. The duality descriptions, namely the simulation particle and
distribution function, are utilized to describe the photon.  By
accurately recovering the governing equations of the unified gas-kinetic
scheme (UGKS), the UGKWP preserves the multiscale dynamics of photon
transport from optically thin to optically thick regime. In the
optically thin regime, the UGKWP becomes a Monte Carlo type particle
tracking method, while in the optically thick regime, the UGKWP
becomes a diffusion equation solver. The local photon dynamics of the
UGKWP, as well as the proportion of wave-described and
particle-described photons are automatically adapted according to the
numerical resolution and transport regime. Compared to the $S_n$-type UGKS,
the UGKWP requires less memory cost and does not suffer ray effect.
Compared to the implicit Monte Carlo (IMC) method, the statistical
noise of UGKWP is greatly reduced and computational efficiency is
significantly improved in the optically thick regime. Several
numerical examples covering all transport regimes from the optically
thin to optically thick are computed to validate the accuracy and
efficiency of the UGKWP method. In comparison to the $S_n$-type UGKS
and IMC method, the UGKWP method may have  several-order-of-magnitude
reduction in computational cost and memory requirement in solving some
multsicale transport problems.
\end{abstract}

\end{frontmatter}

\section{Introduction}
The radiative transfer equation describes photon propagation in the
background medium and has important applications in the fields of
astrophysics \cite{davis2012}, atmospheric physics
\cite{marshak20053d}, optical imaging \cite{klose2002optical} and so
on. In this paper, we focus on the gray radiative transfer equation
with isotropic scattering, which reads
\begin{equation}\label{eq:rt-3D}
  \dfrac{1}{c}\pd{I}{t} + \bsOmega
  \cdot \nabla I = \sigma_s \left(\dfrac{1}{4\pi}\int_{\bbS^2}
  I\dd\bsOmega - I\right) - \sigma_a I + G,
\end{equation}
where $I(t,\bx, \bsOmega)$ is the specific intensity in
time $t$, space $\bx \in \bbR^3$, and angle $\bsOmega$. Here $c$ is
the speed of light, $\sigma_s$ is the scattering coefficient,
$\sigma_a$ is the absorption coefficient, and $G$ is an internal
source of photons.

There are typically two categories of numerical methods for solving
the radiative transfer equations. The first category consists of the
deterministic methods with different ways of discretizing and
modeling, such as the discrete ordinate method (DOM)
\cite{hunter2013comparison, coelho2014advances, chen2015,
roos2016conservation} and the moment methods \cite{frank2006partial,
carrillo2008numerical, vikas2013radiation,
alldredge2016approximating}. The second category is the
stochastic methods, for example, the Monte Carlo method
\cite{fleck1971, lucy1999computing, hayakawa2007coupled}. The Monte
Carlo method is a very popular method for solving the radiative
transfer problems. In comparison with the deterministic methods, it is
more efficient in optically thin regime especially for the
multidimensional cases, and it does not suffer from the ray effect.
However, as a particle method, it unavoidably has statistical noise.
Also,  it becomes inefficient when it comes to diffusive regime. In
diffusive regime where the mean free path is small, photons may go
through a huge number of scatterings during their lifetimes. Direct
simulation of each scattering process for all particles makes the
Monte Carlo method very expensive for capturing the diffusive solutions.
On the other hand, in the diffusive regime the photon transport
process could be well described by the diffusion equation, which could
be solved efficiently. Based on this observation, many hybrid methods
have been developed in order to improve the overall efficiency in
different regimes \cite{fleck1984random, giorla1987random,
densmore2007hybrid, densmore2012hybrid}, where the Monte Carlo method
is used in the optically thin regions and the diffusion equation is
applied to the optically thick regions.  However, as far as we know,
there is still no unified principle for accurate domain decomposition
for different regimes.

Another approach towards releasing the stiffness issue in the
diffusive regime is to develop asymptotic-preserving (AP) DOM
\cite{klar1998asymptotic, naldi1998numerical, jin1999efficient,
  jin2000uniformly, mieussens2013asymptotic, sun2015asymptotic1,
sun2015asymptotic, sun2017multidimensional, sun2017implicit,
sun2018asymptotic} .  One of the examples is the unified gas-kinetic
scheme (UGKS), which couples the particles' transport and collision
process using a multiscale flux function obtained from the integral
solution of the kinetic model equation. The cell size and time step
are not restricted by the mean free path and mean collision time. It
was developed initially in the field of rarefied gas dynamics
\cite{xu2010unified, xu2015direct,jiang} and has been applied to the field
of radiative transfer \cite{mieussens2013asymptotic,
sun2015asymptotic1, sun2015asymptotic, sun2017multidimensional,
sun2017implicit, sun2018asymptotic}, plasma transport
\cite{liu2017unified}, and disperse multi-phase flow
\cite{LIU2019264}. Since it is a discrete ordinate based numerical
scheme, it has no statistical noise, but unavoidably suffers from the
ray effect.

The UGKS provides a general methodology to construct multiscale simulation methods for the transport equations
\cite{xu2015direct,xu-liu-pof}. It consists of two governing equations for the microscopic distribution function and macroscopic
flow variables on the mesh size and time step scales, and a multiscale evolution solution for the interface distribution function.
The time-dependent evolution solution at a cell interface covers the dynamics from the particle free transport
to the hydrodynamic wave interaction with the variation of the ratio of the time step to the local particle collision time.
The original UGKS is constructed based on the discrete velocity method (DVM) or DOM formulations \cite{xu2010unified}.
However, the corresponding purely particle version and wave-particle version of UGKS can be constructed as well \cite{liu-zhu-xu,zhu-liu-zhong-xu}.
In this work, we develop a novel unified gas-kinetic wave-particle
(UGKWP) method for the multiscale photon transport, which combines the
advantages of UGKS and the particle method. To
facilitate understanding, we first describe a purely particle-based
unified gas-kinetic particle (UGKP) method. In the UGKP method, the
photons are described by the particle transport and collision, and
this process is controlled by a multiscale integral formulation.  More specifically, the Monte Carlo particle model is used to
discretize the angular direction of the photon's movement. Based on
the particles' transport nature in the discretized physical space,
particles are categorized into two groups.  Given a fixed time step,
the free-stream particles are accurately tracked by following
the trajectories of the simulation particles, while those particles
that get scattered within the given time step are eliminated and
re-sampled according to their macroscopic variables at the new time
level.  The fluxes across a cell interface from different type
particles are taken into account to update cell averaged
macroscopic variables. In such a way, the multiscale process is
  preserved by coupling  particle free streaming and collision.
  Based on UGKP, a more efficient UGKWP method is proposed.
  Instead of representing all photons by
    simulation particles, a proportion of photon is represented by
an analytical distribution function in UGKWP. Therefore, part of the evaluation of the fluxes,
which were computed by particles in the UGKP method, could
be done analytically with significant reduction in computational cost and
statistical noise, especially in the diffusive regime. The multiscale
flux function of the UGKS is precisely preserved in the UGKWP
implementation. In the diffusive regime, the resulting algorithm would
become a standard central difference scheme for the diffusion
equation without any particle and DOM coordinate in the velocity space. 
In the optically thin regime, it gives a particle tracking
method same as  the Monte Carlo method. In the transition regime, the
ratio of the time step over particle collision time determines the
transport dynamics between the above two limits. The UGKWP is not a hybrid method of domain decomposition, 
but is a method to recover the multiscale modeling in UGKS \cite{xu2015direct} with the help of wave and particle in order to
achieve a highest efficiency in the transport simulation in different regimes.  

The rest of this paper is organized as follows. Section
\ref{sec:preliminary} briefly recalls the basic idea of the unified
gas-kinetic scheme (UGKS) for the linear transport equation. Section
\ref{sec:method} presents the UGKP method for linear photon transport. The UGKWP method is
described in Section \ref{sec:ugkwp}, while the extension to
radiation-material coupling system is discussed in Section \ref{sec:coupled}.
In Section \ref{sec:numerics}, numerical tests are presented to
demonstrate the accuracy and efficiency of the UGKWP method. The final
section is the conclusion.

\section{Review of the UGKS for the linear transport equation}
\label{sec:preliminary}
The unified gas-kinetic scheme (UGKS) was initially developed for the
problems in the field of rarefied gas dynamics \cite{xu2010unified,
xu2015direct}, and has also been successfully applied to problems in
radiative transfer under the finite volume framework
\cite{mieussens2013asymptotic, sun2015asymptotic1, sun2015asymptotic,
sun2017multidimensional, sun2017implicit, sun2018asymptotic}. In this
section, we review the basic idea of the UGKS using the example of the
multidimensional linear transport equation in a purely scattering
medium.

Consider
\begin{equation}
  \dfrac{1}{c}\pd{I}{t} + \bsOmega\cdot\nabla I  =
  \sigma \left(\dfrac{1}{4 \pi}\int_{\bbS^2} I\dd\bsOmega - I\right),
\end{equation}
which gives a non-dimensional equation
\begin{equation}\label{eq:rt-nondimensional}
     \dfrac{\epsilon^2}{c} \pd{I}{t} + \epsilon \bsOmega \cdot\nabla I
   = \sigma \left(\dfrac{1}{4\pi} E - I\right),
\end{equation}
where $E = \int_{\bbS^2} I(\bsOmega) \dd\bsOmega$. The
non-dimensionalization is used same as that in \cite{mieussens2013asymptotic}, and
$\epsilon$ is the ratio between the typical mean free path and the
macroscopic length scale.

The UGKS is based on a finite volume framework. Consider a single cell
$m$. Denote $D_m$ as the area covered by this cell, $\bstau_j$ as the
interfaces of its boundary $\partial D_m$, $\bn_j$ as the outward
normal of $\bstau_j$, and $V_m$ as the volume of cell $m$. Define
$$
  I^n_m = \dfrac{1}{V_m}\int_{D_m} I(t^n, \bx, \bsOmega) \dd \bx
$$
to be the averaged specific intensity $I$ over the spatial cell, and
$$
  E^n_m = \dfrac{1}{V_m}\int_{D_m} E(t^n, \bx) \dd \bx
$$
to be the averaged energy density function $E$ over a spatial cell.
Under the finite volume framework, the discretizations of the two fundamental governing equations for
microscopic and macroscopic variables in UGKS are,
\begin{equation}\label{eq:rt-discretize-fvm}
  \dfrac{I^{n+1}_m - I^n_m}{\Delta t} +   \sum\limits_{\bstau_j \in
  \partial D_m} \dfrac{\phi_{\bstau_j}}{V_m} =
  \dfrac{c \sigma}{\epsilon^2}\left(E^{n+1}_m -
  I^{n+1}_m\right),
\end{equation}
and
\begin{equation}\label{eq:rt-macro-fvm}
  \dfrac{E^{n+1}_m - E^n_m}{\Delta t} +
  \sum\limits_{\bstau_j \in \partial D_m} \dfrac{\Phi_{\bstau_j}}{V_m}  = 0,
\end{equation}
where the microscopic and macroscopic flux terms are respectively
\begin{equation}\label{eq:flux-micro}
  \phi_{\bstau_j} = \dfrac{c}{\epsilon \Delta
  t}\int_{t^n}^{t^{n+1}} \int_{\bstau_j} \left(\bsOmega\cdot\bn_j\right) I(t,
  \bx, \bsOmega) \dd l \dd t,
\end{equation}
and
\begin{equation}\label{eq:flux-macro}
  \Phi_{j+\frac12} = \int_{\bbS^2} \phi_{\bstau_j}(\bsOmega) \dd
  \bsOmega.
\end{equation}

Based on the above microscopic and macroscopic governing equations, the key ingredient of the UGKS is the construction of the multiscale
flux function by adopting the integral solution of the kinetic model
equation \eqref{eq:rt-nondimensional}. The integral solution of equation
\eqref{eq:rt-nondimensional} along the characteristic line gives
\begin{equation}\label{eq:integral-solution}
  \begin{split}
    I(t, \bx, \bsOmega) =
    & \ie^{-\frac{c \sigma(t - t^n)}{\epsilon^2}} I\left(t^n, \bx -
      \frac{c}{\epsilon} \bsOmega(t - t^n)\right) \\
    & + \int_{t^n}^t \ie^{-\frac{c \sigma(t-s)}{\epsilon^2}}
      \times \dfrac{c \sigma}{\epsilon^2} \frac{1}{4\pi}
      E\left(s, \bx - \frac{c}{\epsilon} \bsOmega (t - s)\right) \dd s,
\end{split}
\end{equation}
which is used to construct the numerical fluxes in equation
\eqref{eq:rt-discretize-fvm} and \eqref{eq:rt-macro-fvm}. The integral solution couples transport
with particle collisions, and bridges the kinetic and the hydrodynamic
scale dynamics.
Note that the dynamics in the evolution equations of UGKS through equations \eqref{eq:rt-discretize-fvm}, \eqref{eq:rt-macro-fvm}, and \eqref{eq:integral-solution},
is related to time step $\Delta t$. The ratio of the time step to the local particle collision time determines the
transport regime of the UGKS.

The  numerical flux for microscopic and macroscopic variable updates
is based on the piecewise linear initial reconstruction of $I$ and
$E$ at the beginning of each time step. The discretization of $I$ is based on the DOM in the original UGKS.
It has been proved in \cite{mieussens2013asymptotic} that when $\sigma$
equals $0$, the UGKS tends to a finite volume scheme which is
consistent with free transport solution. More specifically, for 1D
case with uniform grid, the scheme becomes
\begin{displaymath}
  \dfrac{I^{n+1}_j - I^n_j}{\Delta t} + \dfrac{1}{\Delta
  x}\dfrac{c}{\epsilon} \left(\bsOmega \cdot \bn\right) \left(\left(I^n_j
      1_{\bsOmega\cdot \bn >
    0} + I^n_{j+1} 1_{\bsOmega\cdot\bn < 0}\right) - \left(I^n_{j-1}
      1_{\bsOmega \cdot \bn >0} +
  I^n_j 1_{\bsOmega \cdot \bn < 0}\right)\right) = 0.
\end{displaymath}
In the diffusion
limit, with a uniform mesh the UGKS scheme becomes a standard central
difference method for the limit diffusion equation as $\epsilon$ tends
to $0$. Specifically, for 1D case with uniform grid, in the diffusive limit the scheme is
\begin{displaymath}
  \dfrac{E^{n+1}_j - E^n_j}{\Delta t} - \dfrac{1}{\Delta
  x}\left(\dfrac{c}{3 \sigma_{j+\frac12}}\dfrac{E^{n}_{j+1} -
    E^{n}_j}{\Delta x} - \dfrac{c}{3
    \sigma_{j-\frac12}}\dfrac{E^{n}_j -
  E^{n}_{j-1}}{\Delta x}\right) = 0.
\end{displaymath}
More details on the
asymptotic analysis of the UGKS for the radiative transfer equation can be found in
\cite{mieussens2013asymptotic} and \cite{sun2015asymptotic1}.

Following the methodology of the UGKS, we will construct two particle-based
algorithms with multiscale transport property for recovering transport
physics from the kinetic scale to the hydrodynamic scale, i.e., the
unified gas-kinetic particle (UGKP) method and the unified gas-kinetic
wave-particle (UGKWP) method. The UGKP is a purely particle based method
whose computational cost and statistical noise remain the same in all
flow regimes. The UGKWP method is an improvement of the UGKP method
such that the distribution function is represented by the combination
of simulation particles and an analytical distribution function, which
makes it more efficient and less noisy than the UGKP in the diffusive
regime. In both algorithms, for the kinetic scale particle free
transport, the particle trajectories are tracked precisely; for
those particles suffering collisions, their macroscopic
variables will be updated and used to re-sample these particles subsequently.
The multiscale particle methods for equation \eqref{eq:integral-solution} are
constructed through the tracking and re-sampling of particles with the
help of updated macroscopic variables. We will first introduce the
UGKP method in the next section.

\section{The unified gas-kinetic particle (UGKP) method}
\label{sec:method}
In this section, we present a unified gas-kinetic particle (UGKP)
method. Following the direct modeling methodology of UGKS, the
evolution of microscopic simulation particle is coupled with the
evolution of macroscopic energy in UGKP method. The method presented in this
and the next section is based on a single linear transport equation,
\begin{equation}\label{single-bgk}
  \frac{\epsilon^2}{c}\frac{\partial
  I}{\partial t}+\epsilon \bsOmega \cdot
  \nabla I=\sigma\left(\frac{1}{4\pi}E-I\right).
\end{equation}
In the following two subsections, we will introduce the
multiscale evolution equations of particle and macroscopic energy.
For simplicity, the method will be presented for the two-dimensional
case on uniform Cartesian mesh. Its extension to
non-uniform mesh and 3D is
straightforward.

\subsection{Multiscale Particle evolution}
\begin{figure}[H]
  \centering
  \includegraphics[width=0.8\textwidth]{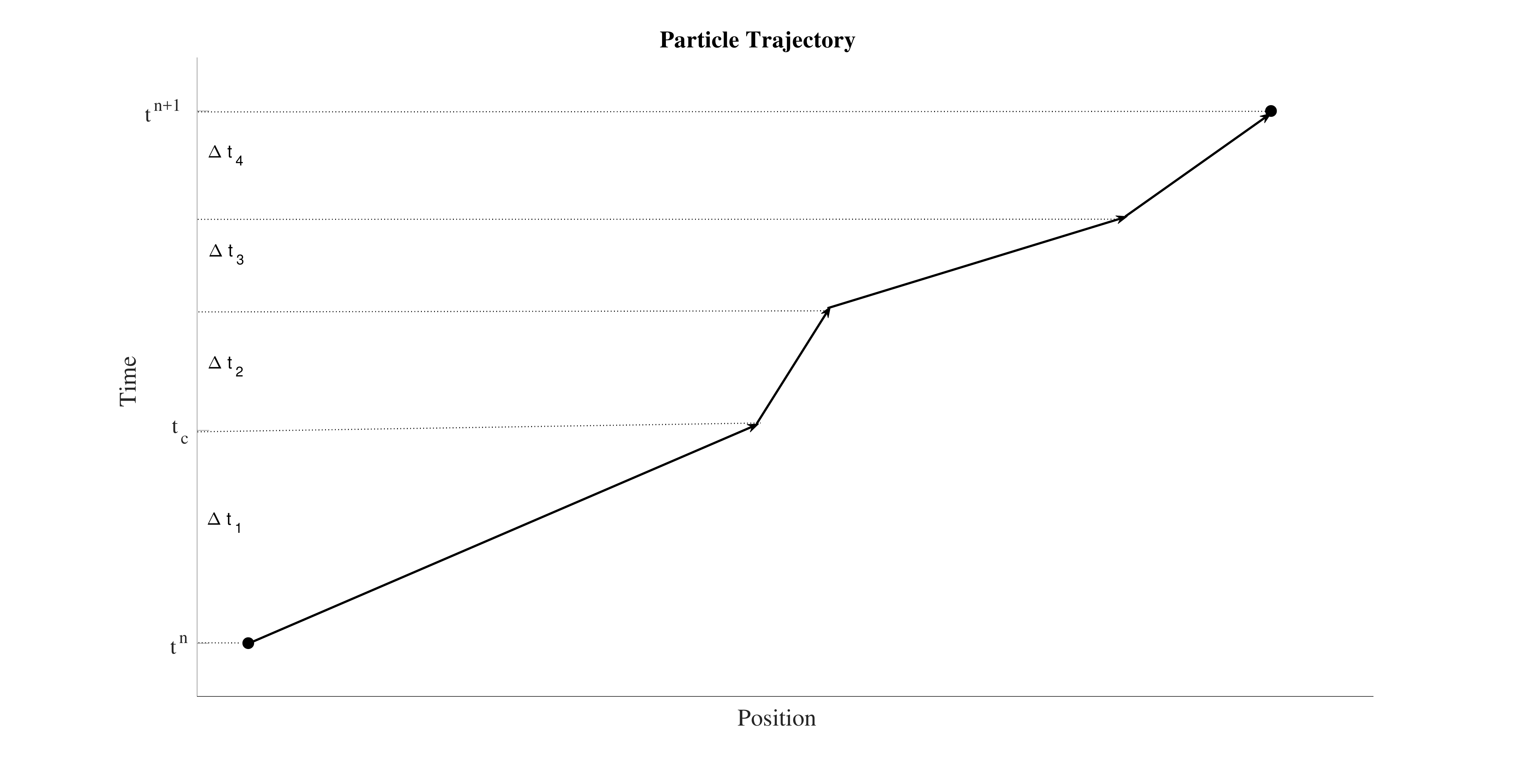}\\
  \caption{Particle trajectory in Monte Carlo method for $t^{n}<t\le
  t^{n+1}$ with sub time steps $\Delta t_1$, $\Delta t_2$, $\Delta
  t_3$ and $\Delta t_4$.}
  \label{trajectory-mc}
\end{figure}

\begin{figure}[H]
  \centering
  \includegraphics[width=0.8\textwidth]{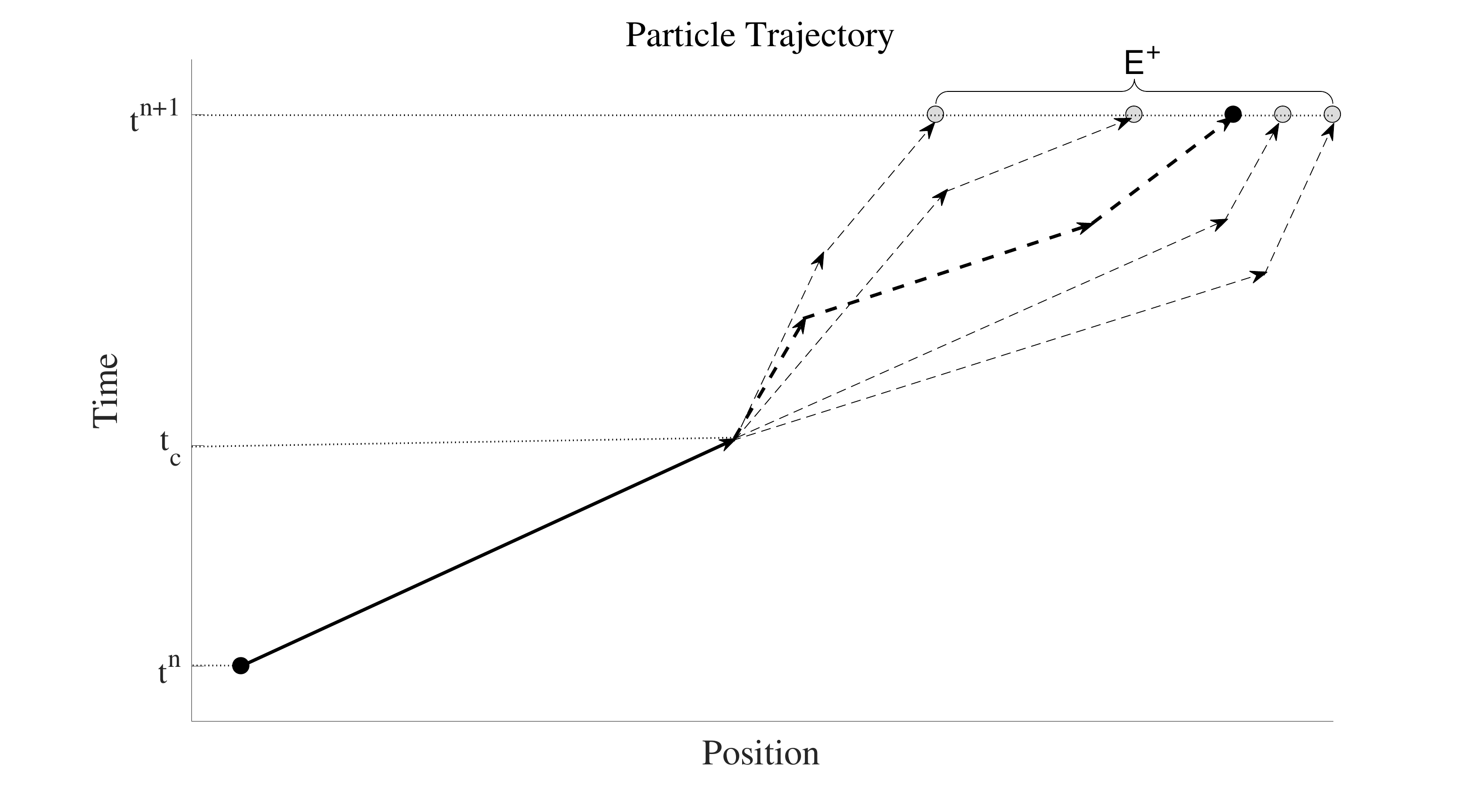}\\
  \caption{UGKP's treatment of particle evolution for $t^{n}<t\le
  t^{n+1}$.}
  \label{trajectory-ugkp}
\end{figure}

The evolution of the simulation particle follows the integral solution
of the linear transport equation \eqref{single-bgk},
\begin{equation}\label{integral-solution}
  I(t,\bx,\bsOmega)=\int_0^{t}e^{-\frac{c\sigma}{\epsilon^2}(t-s)}
  \frac{c\sigma}{\epsilon^2}\frac{1}{4\pi}
  E\left(s,\bx-\frac{c}{\epsilon}\bsOmega(t-s), \bsOmega\right)ds
  +e^{-\frac{c\sigma}{\epsilon^2}t}I_0\left(\bx-\frac{c}{\epsilon}
  \bsOmega t\right).
\end{equation}
The local equilibrium $E$ can be expanded as
\begin{equation}\label{expansion}
  E\left(s,\bx-\frac{c}{\epsilon}\bsOmega(t-s), \bsOmega
  \right)=E(t,\bx, \bsOmega)
  +E_\bx(t,\bx, \bsOmega)\frac{c}{\epsilon}\bsOmega(s-t)+E_t(t,\bx,
  \bsOmega)(s-t)
  +O(\Delta \bx^2,\Delta t^2),
\end{equation}
and the integral solution can be written as
\begin{equation}\label{microevolution}
  I(t,\bx,\bsOmega)=\left(1-e^{-\frac{c\sigma}{\epsilon^2}t}\right)\frac{1}{4\pi}E^e(t,\bx,
  \bsOmega)
  +e^{-\frac{c\sigma}{\epsilon^2}t}I_0(\bx-\frac{c}\epsilon\bsOmega t).
\end{equation}
The first order approximation gives
\begin{equation}\label{1st}
 E^e(t,\bx, \bsOmega)=E(t,\bx, \bsOmega),
\end{equation}
and the second order approximation gives
\begin{equation}\label{2st}
  E^e(t,\bx,\bsOmega)=E(t,\bx, \bsOmega)+\frac{e^{-\frac{c\sigma}{\epsilon^2}t}\left(t+\frac{\epsilon^2}{c\sigma}\right)
  -\frac{c\sigma}{\epsilon^2}}{1-e^{-\frac{c\sigma}{\epsilon^2}t}}\left(E_t(t,\bx,
    \bsOmega)+\frac{c}{\epsilon}\Omega
  E_\bx(t,\bx, \bsOmega)\right).
\end{equation}
The reformulated integral solution Eq.\eqref{microevolution} is the
multiscale governing equation for particle evolution in UGKP method,
which implies
that the simulation particle has a probability of
$e^{-\frac{c\sigma}{\epsilon^2}t}$ to free stream and has a
probability of $1-e^{-\frac{c\sigma}{\epsilon^2}t}$ to get scattered.
The scattered particle follows a velocity distribution $E^e(t,x,
\bsOmega)$, and the first order approximation Eq.\eqref{1st} is used
in this paper. Different from the traditional Monte Carlo method which tracks the
trajectory of each particle as shown in Fig. \ref{trajectory-mc},
UGKP tracks particle trajectory until its first scattering time, and then evolves the
velocity distribution function. 
The scattered particles will be temporarily removed and then get
re-sampled at $t^{n+1}$ from their evolved distribution, as shown in
Fig. \ref{trajectory-ugkp}. The free-streaming particles are named as
the $E_f$ particles, and the scattered particles are named as the
$E^+$ particles. The total energy of the $E^+$ particles is called
$E^+$ energy. In UGKP we do not need to resolve each scattering
process by dividing every time step into several sub time steps and
therefore achieve a high efficiency compared to the traditional Monte
Carlo method. 

The first
scattering time $t_c$ is defined as the time when the first scattering happens to
the particle.
According to Eq.\eqref{microevolution}, the time $t_c$ can be obtained for each particle as 
$$t_c=-\frac{\epsilon^2}{c\sigma}\ln \xi ,$$
where $\xi$ is a random number which follows a uniform distribution on $(0,1)$.
Simulation particles can be categorized into two groups based on
$t_c$. The first group of simulation particle satisfies 
$$t_{c}\ge \Delta t = t^{n+1}-t^n $$ 
and does not suffer any collision during a time
step $\Delta t$, and this particle position is updated by
\begin{displaymath}
  \bx^{n+1}=\bx^{n}+\dfrac{c \bsOmega^{n}}{\epsilon}\Delta t.
\end{displaymath}
The second group of particle satisfies $t_{c}< \Delta t $, those
particles suffer one or multiple times of scattering during $[t_c,
t^{n+1}]$, with the assumption $t^n = 0 $ for simplification. 
Therefore, these particles will only be accurately tracked
for $t\in[t^n,
t_c)$ and be re-sampled from
$E^+(t^{n+1}, \bx,
\bsOmega)$ at $t^{n+1}$.
$E^+(t^{n+1}, \bx, \bsOmega)$ is obtained from the evolved macroscopic
energy. In the UGKP method, the evolution of particle and macroscopic
energy are closely coupled. The multiscale evolution equation of macroscopic energy is given in the next subsection.

\subsection{Multiscale macroscopic energy evolution}
\begin{figure}
  \centering
  \includegraphics[width=0.8\textwidth]{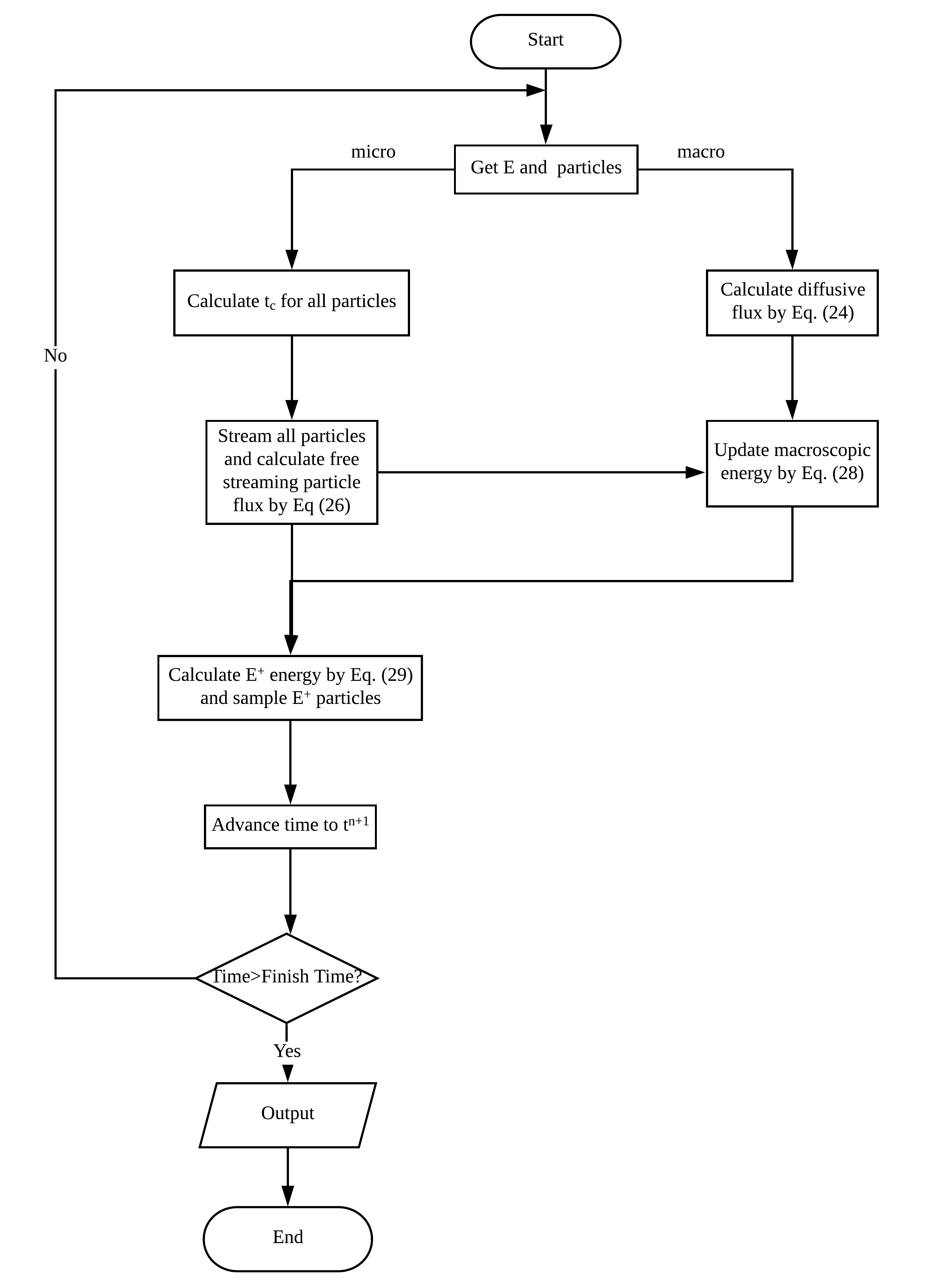}\\
  \caption{Flowchart of UGKP.}
  \label{flowchart1}
\end{figure}

The update of the
macroscopic variable follows the finite volume framework
\begin{equation} \label{eq:macro-update}
  E_{m}^{n+1} = E_m^n - \dfrac{\Delta t}{V_m} \sum\limits_{\bstau_j
  \in \partial D_m} \Phi_{E, \bstau_j} - \dfrac{1}{V_m}
  \sum\limits_{\bstau_j \in \partial D_m} W_{\bstau_j},
\end{equation}
where the flux $\Phi_E$ is  the diffusive flux and $W$ is the
contribution of the free streaming flux.
Based on \eqref{microevolution}, the diffusive flux is computed same as that in the UGKS,
\begin{displaymath}
  \Phi_{E, \bstau_j} = \dfrac{c}{\epsilon \Delta t}
  \int_{t^n}^{t^{n+1}} \int_{t^n}^t \int_{\bstau_j} \int_{\bbS^2} \left(\bsOmega \cdot
  \bn_j\right) \dfrac{c \sigma_{\bstau_j}}{\epsilon^2} \ie^{-\frac{c
  \sigma_{\bstau_j}(t-s)}{\epsilon^2}} \dfrac{1}{4\pi} E\left(s,
  \bx - \dfrac{c}{\epsilon}(t - s) \bsOmega\right)
  \dd\bsOmega \dd l \dd s \dd t.
\end{displaymath}
We adopt the piecewise linear reconstruction for $E$,
\begin{equation}\label{eq:E-reconstruct}
  \begin{split}
    E\left(s, \bx_{\bstau_j} - \frac{c}{\epsilon} (t - s)
    \bsOmega\right) = & E(t^n,
  \bx_{\bstau_j}) + \pd{E}{t}(t^n, \bx_{\bstau_j}) \times (s -
  t_{n}) \\
  &
  -\frac{c}{\epsilon}(t - s) \bsOmega \cdot \nabla E(t,
  \bx_{\bstau_j}),
\end{split}
\end{equation}
and the explicit central difference discretization of $\nabla E$,
\begin{displaymath}
  \nabla E(t^n, \bx_{\bstau_j})
  \approx \dfrac{E^n_m - E^n_k}{\bx_m - \bx_k},
\end{displaymath}
where $\bx_m$, $\bx_k$ are the barycenters
of cells $m$ and $k$ with the same interface
$\bstau_j$.
Then the diffusive flux is given by
\begin{equation}\label{eq:diffusive-flux}
  \Phi_{E, \bstau_j} = l_j \alpha_{\bstau_j} \bn_j \cdot \left(\dfrac{E^n_m - E^n_k}{\bx_m -
  \bx_k}\right),
\end{equation}
where
\begin{displaymath}
  \alpha_{\bstau_j} = -\frac{c}{3 \sigma_{\bstau_j}}
  \left(-\dfrac{2}{\left(\frac{c \sigma_{\bstau_j} \Delta t}{\epsilon^2}\right)}\left(1
    - \ie^{-\frac{c \sigma_{\bstau_j} \Delta t}{\epsilon^2}} \right) +
  \left(1 + \ie^{-\frac{c \sigma_{\bstau_j} \Delta
t}{\epsilon^2}}\right)\right).
\end{displaymath}
The free transport flux $W$ is contributed by the particle free streaming,
which is computed by
\begin{equation}\label{eq:free-stream}
  W_{\bstau_j} = \sum_{k\in S(\bstau_j)} \text{sign}(\bn_j \cdot
  \bsOmega_k) \omega_k,
\end{equation}
where $\omega_k$ and $\bsOmega_k$ are respectively the weight and
velocity of the particle $k$, and $S(\bstau_j)$ is the
set of all particles which are transported
across interface $\bstau_j$ during the free streaming process.
It recovers the other part of UGKS flux
\begin{equation}\label{eq:macro-flux-free-stream}
  \Phi_{f,\bstau_j} = \dfrac{c}{\epsilon \Delta t}
  \int_{t^n}^{t^{n+1}} \int_{\bstau_j} \int_{\bbS^2} \left(\bsOmega \cdot \bn_j\right)
  \ie^{-\frac{c \sigma_{\bstau_j}(t - t^n)}{\epsilon^2}} I\left(t^n,
    \bx - \dfrac{c}{\epsilon}(t - t^n)\bsOmega,
  \bsOmega\right) \dd\bsOmega \dd l \dd t.
\end{equation}
With the diffusive flux Eq. \eqref{eq:diffusive-flux} and the free
transport flux Eq. \eqref{eq:free-stream}, the macroscopic energy $E^{n+1}$ can be updated by
\begin{equation}\label{eq:macro-eq}
  E^{n+1}_{m} = E^n_{m} - \dfrac{\Delta t}{V_m}\sum\limits_{\bstau_j \in
  \partial D_m} l_j \alpha_{\bstau_j} \bn_j \cdot \left(\dfrac{E^n_m - E^n_k}{\bx_m -
  \bx_k}\right) - \dfrac{1}{V_m}\sum\limits_{\bstau_j \in \partial
  D_m} W_{\bstau_j}.
\end{equation}
With the updated $E^{n+1}$, the energy $E^+$ of the scattered particles can be obtained and be used to
re-generate particles,
\begin{equation}\label{eq:Ep-ugkp}
  E^+_{m} = E^{n+1}_m - \dfrac{1}{V_m} \sum_{k \in Q(m)} \omega_k,
\end{equation}
where $Q(m)$ is the set of $E_f$ particles in cell $m$ at $t^{n+1}$.
In summary, the evolution of particle provides the free-stream flux
for macroscopic energy evolution, and the updated energy $E^+$  provides the
distribution for the corresponding particles. The evolution of microscopic and
macroscopic quantities are closely coupled, and the algorithm of UGKP
is shown in Fig. \ref{flowchart1}.
\begin{figure}
  \centering
  \includegraphics[width=1.0\textwidth]{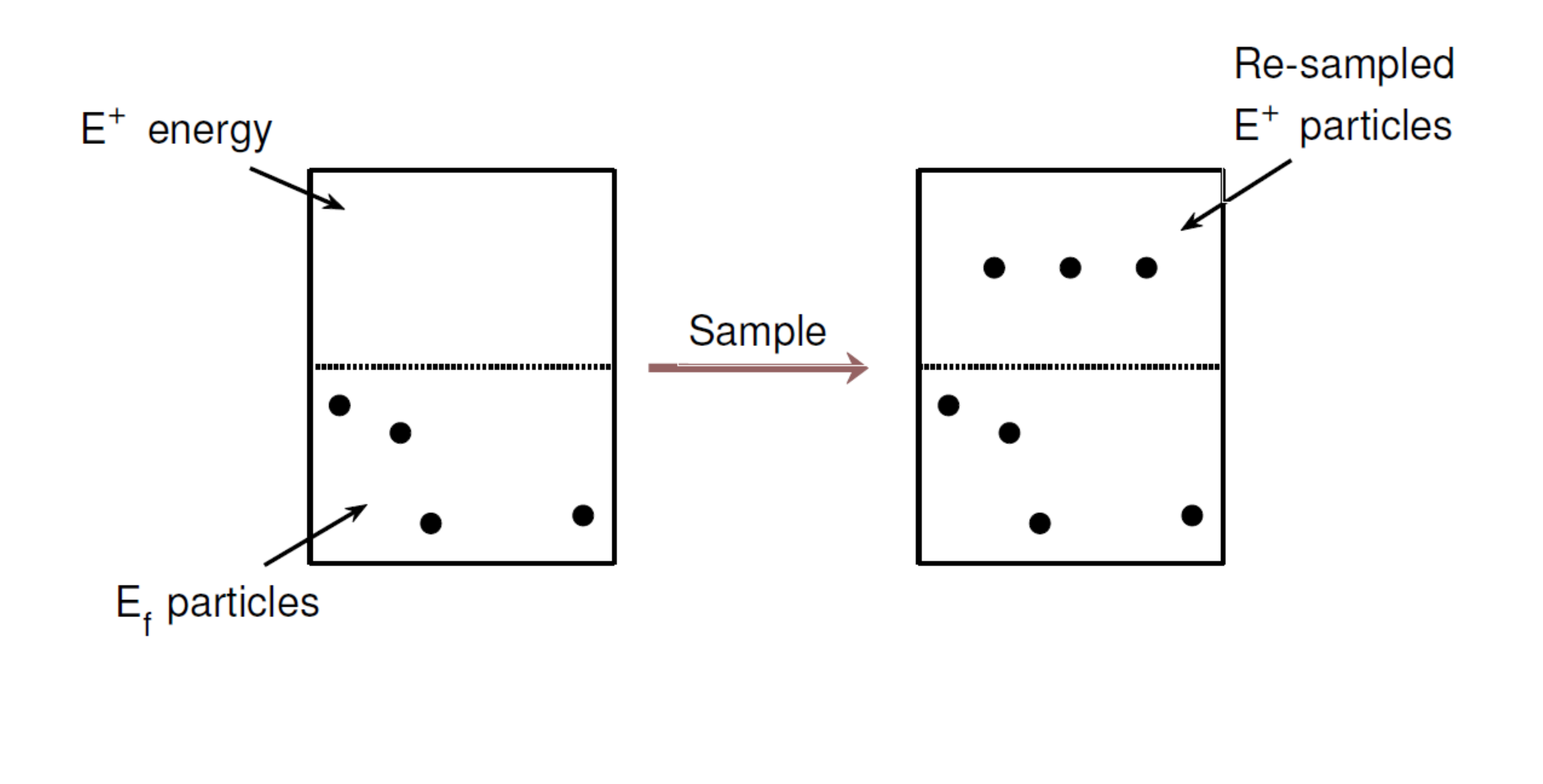}\\
  \caption{UGKP re-sampling process at $t=t^{n+1}$.}
  \label{sample-ugkp}
\end{figure}

\section{Unified gas-kinetic wave-particle (UGKWP) method}\label{sec:ugkwp}

\subsection{The multiscale evolution of particles and macroscopic
variable}
In UGKP, the $E^+$ particles are re-sampled from an equilibrium
distribution at $t=t^{n+1}$.
 According to the
integral solution \eqref{integral-solution}, $E^+$ particles have a
probability $1 - \ie^{-\frac{c \sigma \Delta
t}{\epsilon^2}}$ to get scattered in the next time
step from $t^{n+1}$ to $t^{n+2}$. We define the scattered particles
in $t \in [t^{n+1}, t^{n+2}] $  as $E^+_W$ particles and the
non-scattering particles as $E^+_P$ particles. As the
$E^+_W$ particles will be deleted after $t_c$ within $t \in [t^{n+1},
t^{n+2}] $,  and their
contribution to the free transport flux could be
computed analytically, only the $E^+_P$ particles need to be
re-sampled and they keep on free transport in the step from $t^{n+1}$
to $t^{n+2}$.
For the $E^+_W$ particles, we only need to store the
distribution function instead of re-sampling particles from it. Based on this
observation, we propose a more efficient unified gas-kinetic
wave-particle (UGKWP) method.

\begin{figure}
  \centering
  \includegraphics[width=1.0\textwidth]{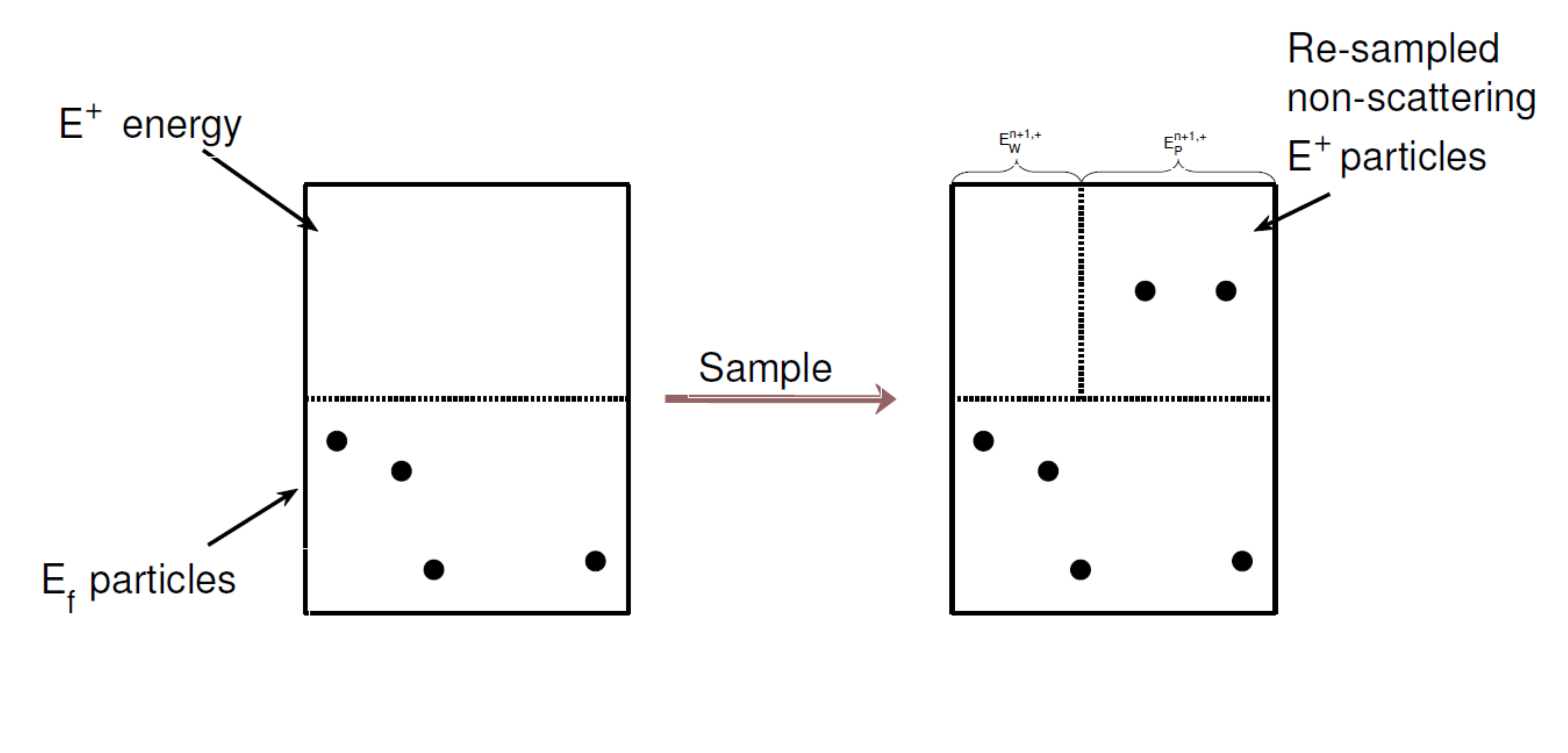}\\
  \caption{UGKWP re-sampling process at $t = t^{n+1}$.}
  \label{sample-ugkwp}
\end{figure}

The unified gas-kinetic wave-particle(UGKWP) method improves UGKP in two aspects:
\begin{enumerate}
  \item Only $E^+_P$ particles are sampled as shown in Fig.
    \ref{sample-ugkwp};
  \item The free transport flux contributed by $E^+_W$ can be  calculated analytically.
\end{enumerate}
The finite volume energy evolution follows
\begin{equation} \label{eq:plus}
  E^{n+1}_m=E^n_m-\frac{\Delta t}{V_m}\sum_{\bstau_j\in\partial
  D_m}(\Phi_{E,\tau_j}+\Phi_{fw,\bstau_j}) - \frac{1}{V_m} \sum_{\bstau_j
  \in \partial D_m} W_{\bstau_j},
\end{equation}
where the diffusive flux $\Phi_{E,\bstau_j}$ is computed using Eq.
\eqref{eq:diffusive-flux}, and the free transport flux contributed by $E^+_W$ is calculated by
\begin{displaymath}
  \begin{split}
    \Phi_{fw,\bstau_j}= & \frac{c}{\epsilon \Delta
  t}\int_{t^n}^{t^{n+1}}\int_{\bstau_j}\int_{\mathcal{S}^2}(\Omega\cdot
\bn_j)e^{-\frac{c\sigma_{\bstau_j}(t-t^n)}{\epsilon^2}}E^+\left(t^n,x-\frac{c}{\epsilon}\left(t-t^n\right)\bsOmega,\bsOmega\right)d\bsOmega
  dldt \\
  & - \frac{c}{\epsilon \Delta
  t}\int_{t^n}^{t^{n+1}}\int_{\bstau_j}\int_{\mathcal{S}^2}(\bsOmega\cdot
\bn_j) E^+_P
\left(t^n,x-\frac{c}{\epsilon}\left(t-t^n\right)\bsOmega,\bsOmega\right)d\bsOmega
  dldt.
\end{split}
\end{displaymath}
The free transport flux $W_{\bstau_j}$ contributed from the
particles in Eq. \eqref{eq:plus} is computed by particle tracking,
\begin{equation}\label{eq:free-stream-flux-ugkwp}
  W_{\bstau_j} = \sum_{k \in S(\bstau_j)} \text{sign}(\bn_j \cdot
  \bsOmega_k) \omega_k,
\end{equation}
where $S(\bstau_j)$ is the set of all particles which are
transported across the cell interface $\bstau_j$ during the free
streaming process. In the particle free transport process, the re-sampled $E^+_P$ particles will always have $t_c > \Delta t$
and keep free transport without collision in the whole time step.
The other particles from $E_f$ will transport according to their individual generated $t_c$.
With the solution of $E^{n+1}$, Eq.\eqref{eq:Ep-ugkp} is used to get $E^+$, which is divided into 
\begin{displaymath}
  E^+_P = \ie^{-\frac{c\sigma \Delta t}{\epsilon^2}} E^+,
\end{displaymath}
and
\begin{displaymath}
  E^+_W = E^+ - E^+_P,
\end{displaymath}
and only particles from $E^+_P$ will be generated and keep on free transport in the whole next time step.

\begin{figure}
  \centering
  \includegraphics[width=0.8\textwidth]{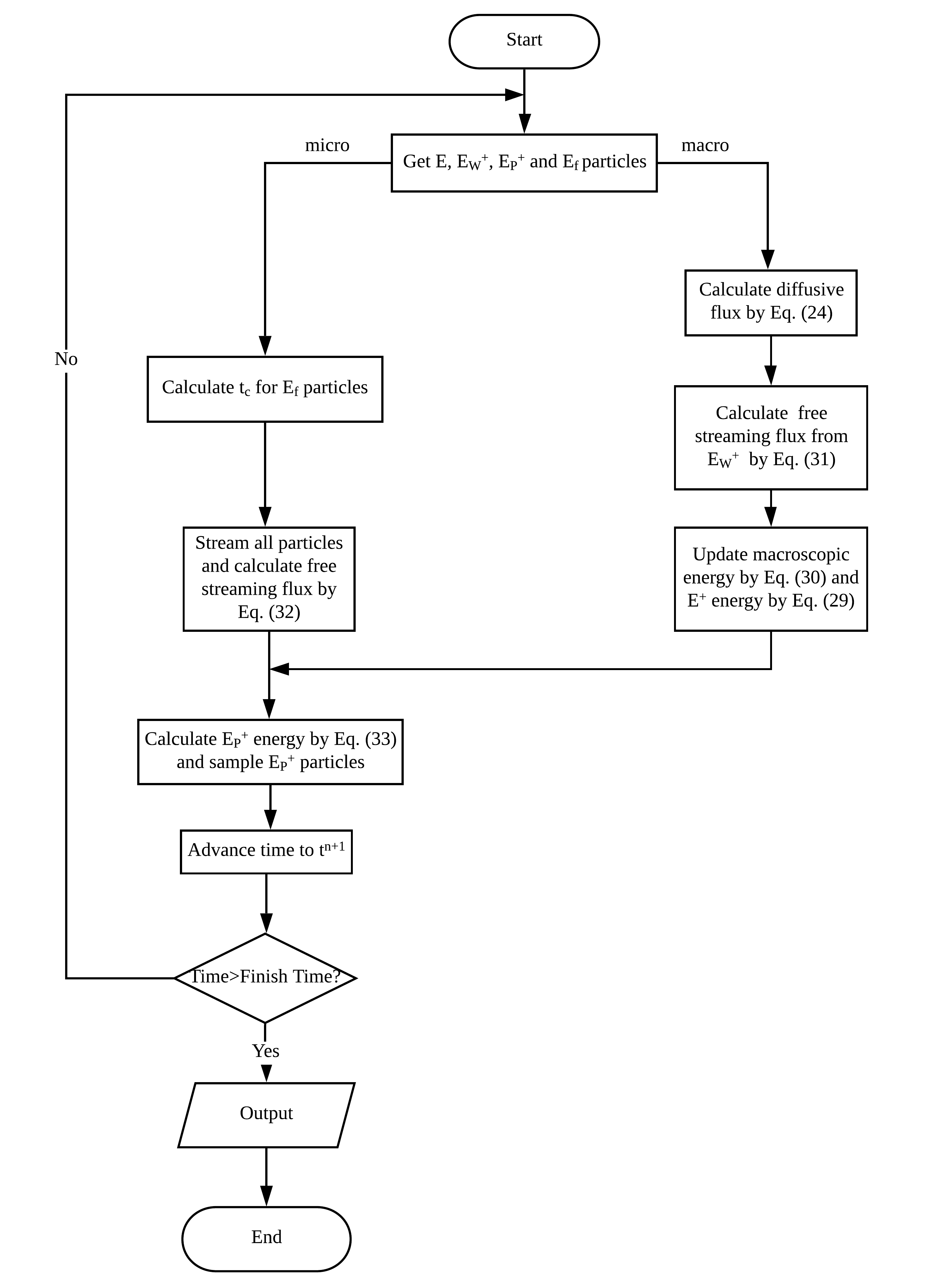}\\
  \caption{Flowchart of UGKWP.}
  \label{flowchart2}
\end{figure}

The algorithm of UGKWP is shown in Fig. \ref{flowchart2}.

\subsection{Properties of the UGKWP algorithm}
The unified gas-kinetic wave-particle method satisfies the following properties:
\begin{enumerate}
  \item The microscopic and macroscopic evolutions are consistently evaluated based on the wave-particle decomposition.
    The macroscopic energy
    density $E$ is the sum of the energy from the un-scattering particles and the scattering one $E^+$.
    The evolution of $E^+_W$ particles can be evaluated analytically.
  \item In the diffusive limit, $\epsilon \rightarrow 0$ and
    $\ie^{-\frac{\sigma_{\bstau_j} \Delta t}{\epsilon^2}} \rightarrow
    0$, essentially no particle can be freely transported and survived within a time step.
    Therefore, no particle is re-sampled from the $E^+$ energy. The algorithm becomes a time-explicit central difference
    solver of the diffusion equation. For instance, in 1D case with
    uniform grid, the scheme tends to the following limiting equation
   \begin{displaymath}
      \dfrac{E^{n+1}_j - E^n_j}{\Delta t} - \dfrac{1}{\Delta
      x}\left(\dfrac{c}{3 \sigma_{j+\frac12}}\dfrac{E^{n+1}_{j+1} -
        E^{n+1}_j}{\Delta x} - \dfrac{c}{3
        \sigma_{j-\frac12}}\dfrac{E^{n+1}_j -
      E^{n+1}_{j-1}}{\Delta x}\right) = 0.
    \end{displaymath}
    Note that in the diffusion regime, different from DOM method there is no discrete velocity space discretization in UGKWP.
  \item In the free transport limit, $\sigma \rightarrow 0$ and
    $\ie^{-\frac{\sigma_{\bstau_j} \Delta t}{\epsilon^2}} \rightarrow 1$,
    each particle is traced exactly by free transport with
    probability 1. In this case, the UGKWP could recover the
    exact solution for individual particle.
\end{enumerate}

\section{Extension to the coupled system  of gray radiative transfer and material energy evolution}
\label{sec:coupled}
This section extends UGKWP to solve the coupled system of gray radiative transfer and material
temperature equation,
\begin{equation}\label{eq:rt-coupled}
  \left\{\begin{split}
  & \dfrac{\epsilon^2}{c} \pd{I}{t} + \epsilon\bsOmega \cdot \nabla I
      = \sigma \left(\dfrac{1}{4\pi}
    a c T^4 - I\right) + G, \\
    & \epsilon^2 C_v \pd{T}{t} = \sigma \left(\int_{\bbS^2}
    I(\bsOmega) \dd\bsOmega - a c
  T^4\right).
\end{split}\right.
\end{equation}
Define $u_r = a T^4$ and $\beta = \pd{u_r}{T}$, then the second equation
can be re-written as
\begin{equation}\label{eq:energy}
  \pd{u_r}{t} = C_v^{-1} \beta \dfrac{\sigma}{\epsilon^2}
  \left(\int_{\bbS^2} I(\bsOmega) \dd\bsOmega
  - c u_r\right).
\end{equation}
The implicit Monte Carlo method proposed by Fleck and Cummings in
\cite{fleck1971} has been shown to be an effective technique for
solving non-linear, time-dependent, radiative transfer problems and is
widely used in the radiative transfer community. Fleck's implicit
Monte Carlo method uses an effective scattering process to approximate
the absorption and emission of radiation by the background medium.
This treatment allows it to take larger time steps than that in a
purely explicit method. Here the similar semi-implicit discretization
for material temperature evolution will be employed.  Specifically, Eq.
\eqref{eq:energy} is discretized by
\begin{displaymath}
  \dfrac{u^{n+1}_r - u^n_r}{\Delta t} = C_v^{-1}\beta^n
  \dfrac{\sigma}{\epsilon^2} \left(E
- c u^{n+1}_r\right),
\end{displaymath}
which gives
\begin{equation}\label{eq:discretize-energy}
  u^{n+1}_r = \dfrac{1}{1 + c C_v^{-1}\beta^n \times \frac{\sigma \Delta
  t}{\epsilon^2}} u^n_r +
  \dfrac{C_v^{-1} \beta^n \times \frac{\sigma \Delta t}{\epsilon^2}}{1 + c C_v^{-1} \beta^n
  \times \frac{\sigma \Delta t}{\epsilon^2}} E.
\end{equation}
With the definition
\begin{displaymath}
  \sigma_a = \dfrac{\sigma}{1 + c C_v^{-1} \beta^n \times
  \frac{\sigma \Delta t}{\epsilon^2} },
  \quad \sigma_s = \sigma - \sigma_a,
\end{displaymath}
substituting Eq. \eqref{eq:discretize-energy} into Eq. \eqref{eq:rt-coupled} yields
\begin{displaymath}
  \dfrac{\epsilon^2}{c} \pd{I}{t} + \epsilon\bsOmega \cdot \nabla I =
  \sigma_s\left(\frac1{4\pi}
  E - I \right) + \sigma_a\left(\frac{1}{4 \pi} c u^n_r - I \right) +
  G.
\end{displaymath}
An operator splitting approach is used to solve the above system subsequently by
 the linear kinetic equation
\begin{equation}\label{eq:split-scattering}
  \dfrac{\epsilon^2}c\pd{I}{t} + \epsilon \bsOmega \cdot \nabla I =
  \sigma_s\left(\dfrac1{4\pi} E - I\right) + G,
\end{equation}
 the radiation energy exchange,
\begin{equation}\label{eq:split-exchange}
  \dfrac{\epsilon^2}{c} \pd{I}{t} = \sigma_a \left(\dfrac1{4\pi}
  c u^n_r - I\right),
\end{equation}
and the update of material energy.
Here Eq. \eqref{eq:split-scattering} is solved using the algorithm
introduced in Section \ref{sec:ugkwp}. The exact solution of Eq.
\eqref{eq:split-exchange} is
\begin{equation}\label{eq:exchange-solution}
  I^{n+1} = \exp\left(-\dfrac{c \sigma_a \Delta t}{\epsilon^2}\right)
  I^\ast + \left[1 - \exp\left(-\dfrac{c \sigma_a \Delta
  t}{\epsilon^2}\right)\right] \dfrac{1}{4\pi} c u^n_r,
\end{equation}
where $I^\ast$ is solved from Eq. \eqref{eq:split-scattering} by
UGKWP.  The exponential decay term in Eq. \eqref{eq:exchange-solution}
is implemented by modifying the weight of particles, while the source
term in Eq. \eqref{eq:exchange-solution} is added to $E^+$ energy. Afterwards, the energy change of particles is summed up, and
the
material temperature and $u_r$ are updated by energy conservation.
The above UGKWP can be further improved if the technique in \cite{sun2015asymptotic} is incorporated into the current scheme,
where both macroscopic equations for the radiation energy and material energy are solved iteratively first before updating the
radiation intensity $I$.

\section{Numerical Experiments}
\label{sec:numerics}
In this section, we present six numerical examples to validate the
proposed UGKWP method.
In all these examples, the UGKWP method obtains
results which are in good agreement with that of the Monte Carlo method.  At the same
time, the UGKWP  is more efficient and less noisy compared with
the Monte Carlo method in the near diffusive regime.  All computations
are performed in sequential code on a computer with Intel i7-8700K CPU and $64$ GB
memory. In all simulations, the time step is determined by $\Delta t =
\text{
CFL} \times \epsilon \Delta x / c$, with $\text{CFL} = 0.4$ for 1D examples and
$\text{CFL} = 0.2$ for 2D examples.

\subsection{Inflow into purely scattering homogeneous medium}
We first consider the behaviour of the UGKWP method for
purely scattering homogeneous medium. Tests in this section are for
non-dimensional linear equation
\begin{displaymath}
  \epsilon\pd{I}{t} + \mu \pd{I}{x} =
  \dfrac{\sigma}{\epsilon}\left(\dfrac12 E - I\right),
\end{displaymath}
defined on the semi-infinite spatial domain $x \in [0, \infty)$ with
an isotropic inflow condition imposed on the left boundary. For the
numerical simulation, the spatial domain is taken to be $[0, 1]$. Inflow
boundary condition is imposed at $x = 0$ with the incoming specific
intensity $I(t,0,\mu)= \frac12$.  The initial value is $I(\mu) = 0$
for all $x$.

The results of both the UGKWP method and the Monte Carlo method are
obtained using $200$ grids in space.  As we are targeting to develop a
method that automatically bridges the
optically thin and optically thick regimes, the parameters cover the
rarefied ($\epsilon \gg \Delta x$), the intermediate ($\epsilon
\approx \Delta x$), and the diffusive ($\epsilon \ll \Delta x$)
regimes, as defined in \cite{jin1998diffusive}.  When $\epsilon$ is
small, i.e. in the diffusive regime, the current method can use a much
larger cell size and time step than the particle mean free path and
collision time. Also, due to the exponential factor in re-sampling,
fewer particles are used in UGKWP  than the Monte Carlo
method when
$\epsilon$ is small, and UGKWP becomes much more efficient and less
noisy than the Monte Carlo method in those cases. Table \ref{tab:timecompare-homo}
compares the computational time  of the Monte Carlo method and the UGKWP
method
for different $\epsilon$, while Table \ref{tab:parcompare-homo}
compares the maximum number of particles used. Notice that
no particle is generated in UGKWP method when $\epsilon = 10^{-4}$
and the method only solves a diffusion equation. For $\epsilon = 10^{-4}$, the Monte
Carlo method takes more than half an hour while the UGKWP method takes
only $16$ seconds, showing that the UGKWP method is much more efficient
than the Monte Carlo method in the diffusive regimes.
\begin{table}[htbp]
    \centering
\caption{Comparison of computational time of the Monte Carlo method and
UGKWP for inflow into homogeneous medium under different
parameters.}\label{tab:timecompare-homo}
\setlength{\tabcolsep}{5mm}{
    \begin{tabular}{ccc}
      \toprule
      $\epsilon$  & MC  & UGKWP\\
        \hline
        $1$ & $0.12$s  & $0.14$s\\
        \hline
        $10^{-2}$ & $1.49$s & $1.35$s \\
        \hline
        $10^{-4}$ & $2587$s & $16$s \\
        \bottomrule
      \end{tabular}}
\end{table}
\begin{table}[htbp]
    \centering
\caption{Comparison of the maximum number of particles used by the Monte Carlo method and
UGKWP for inflow into homogeneous medium under different
parameters.}\label{tab:parcompare-homo}
\setlength{\tabcolsep}{5mm}{
    \begin{tabular}{ccc}
      \toprule
      $\epsilon$  & MC  & UGKWP\\
        \hline
        $1$ & $13053$  & $15612$\\
        \hline
        $10^{-2}$  & $8054$ & $6920$ \\
        \hline
        $10^{-4}$ & $8343$ &  $0$ \\
        \bottomrule
      \end{tabular}}
\end{table}

\begin{figure}[htbp]
  \centering
  \subfigure[Comparison of $E$ between
    UGKWP and the Monte Carlo
  solution for $\epsilon = 1$.]{
    \label{fig:homo-kinetic}
  \includegraphics[width=0.32\textwidth]{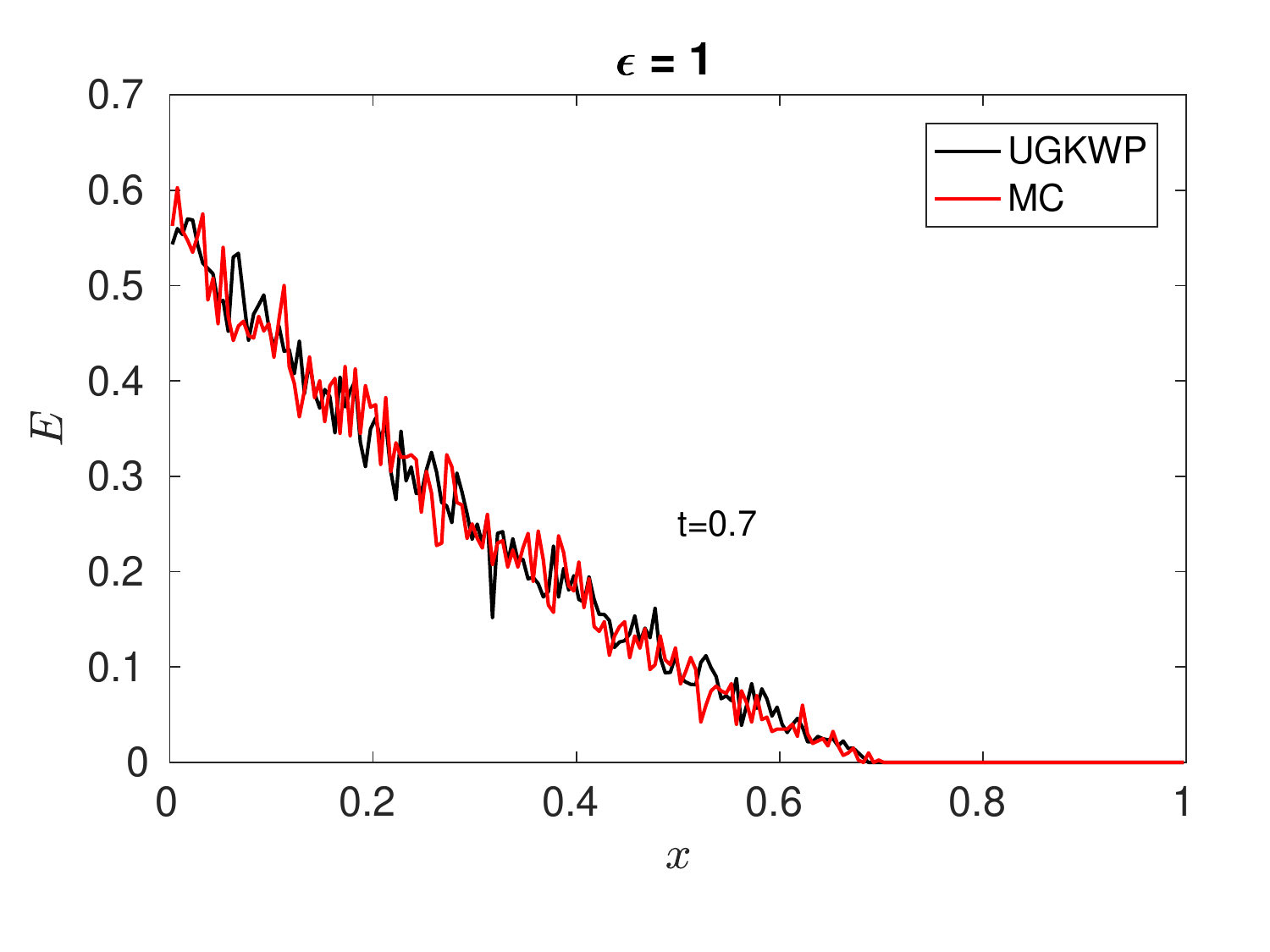}}
  \subfigure[Comparison of $E$ between
  UGKWP and the Monte Carlo solution for $\epsilon = 10^{-2}$.]{\label{fig:homo-intermediate}
  \includegraphics[width=0.32\textwidth]{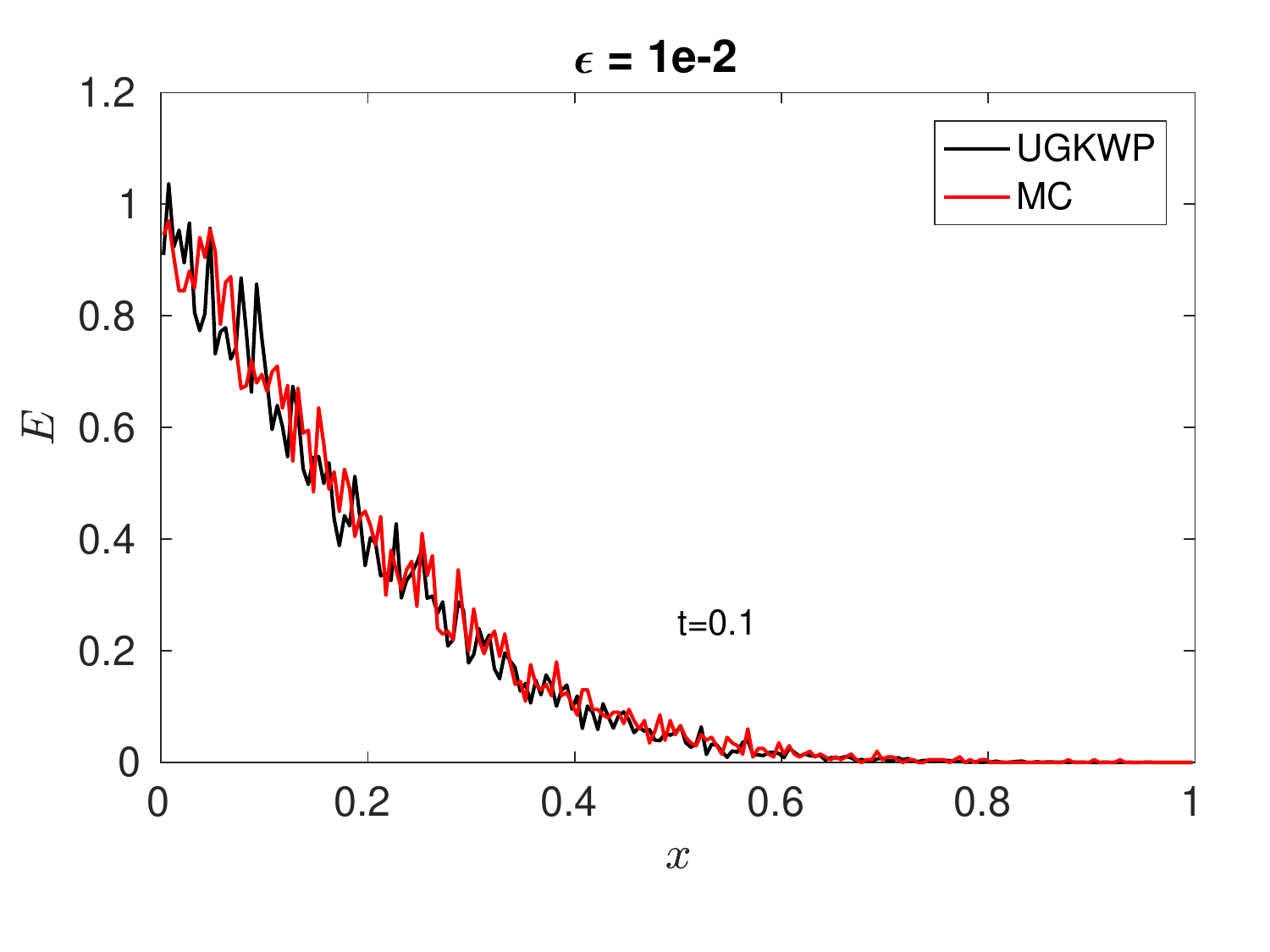}}
  \subfigure[Comparison of $E$ between UGKWP and the Monte Carlo
  solution for $\epsilon = 10^{-4}$.]{\label{fig:homo-diffusive}
  \includegraphics[width=0.32\textwidth]{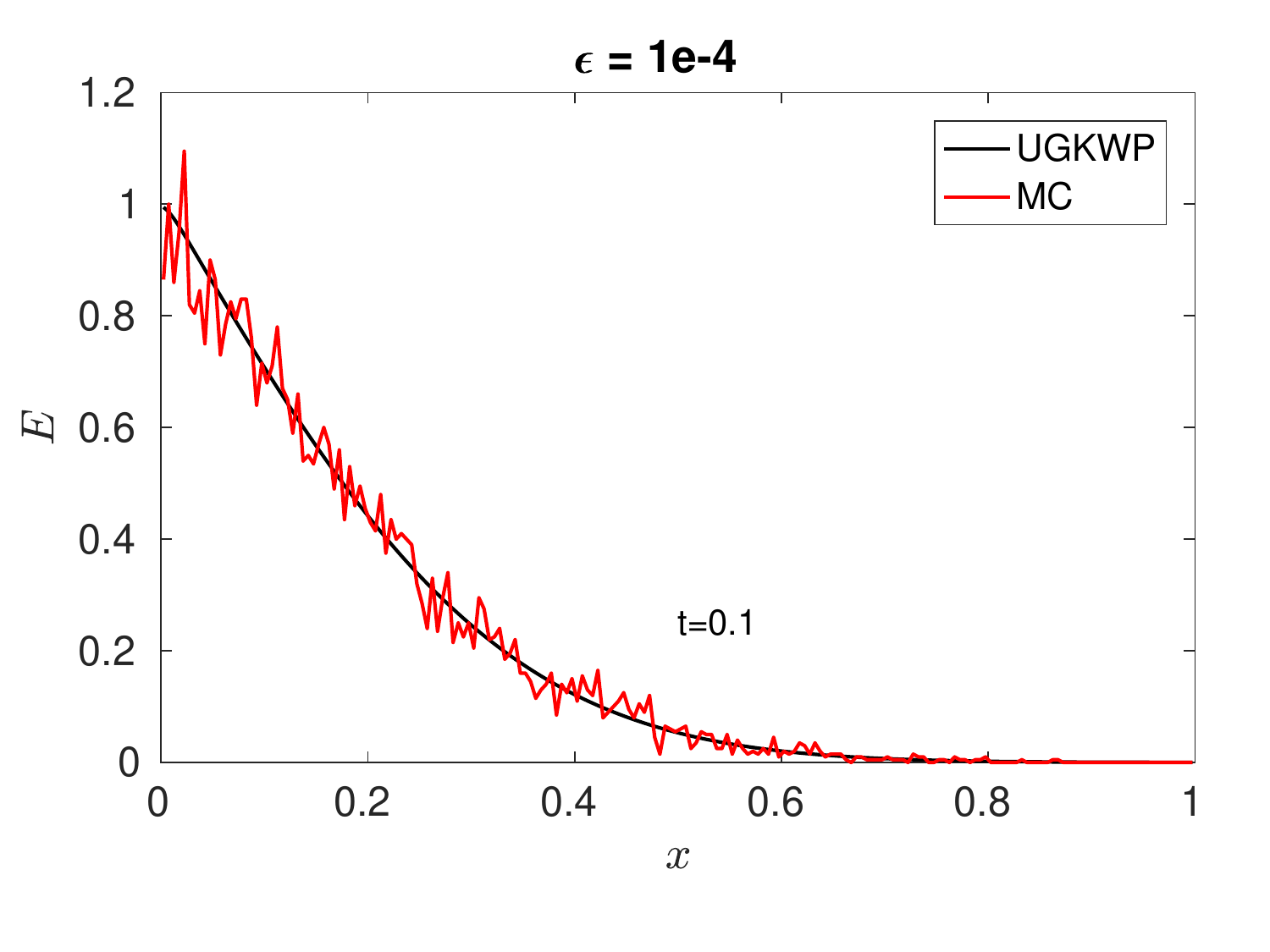}}
  \caption{Numerical results of homogeneous medium.}
  \label{fig:homogeneous}
\end{figure}

In Figure \ref{fig:homogeneous} the numerical results of the UGKWP
method are compared with the solutions of the Monte Carlo method for
different $\epsilon$.
In this example, the UGKWP method gives solutions which
are almost identical with the Monte Carlo solution for all flow
regimes. Also, note that the UGKWP solution is smooth for $\epsilon =
10^{-4}$, while the Monte Carlo solution has statistical noise.

\subsection{Inflow into purely scattering heterogeneous medium}
We next consider the UGKWP method for purely
scattering heterogeneous medium. For this example, the set-up for the
computational domain as well as that of the initial and boundary
conditions are the same as in the previous section. Tests in this
section are for the dimensional linear equation
\begin{displaymath}
  \pd{I}{t} + \mu \pd{I}{x} =
  \sigma \left(\dfrac12 E - I\right).
\end{displaymath}
Both the UGKWP method and the Monte Carlo method use $200$ grids in space.

\begin{figure}[htbp]
  \centering
  \subfigure[Comparison of $E$ between UGKWP and the Monte Carlo
  solution in thick-to-thin heterogeneous medium.]{
    \label{fig:thick2thin-E}
  \includegraphics[width=0.48\textwidth]{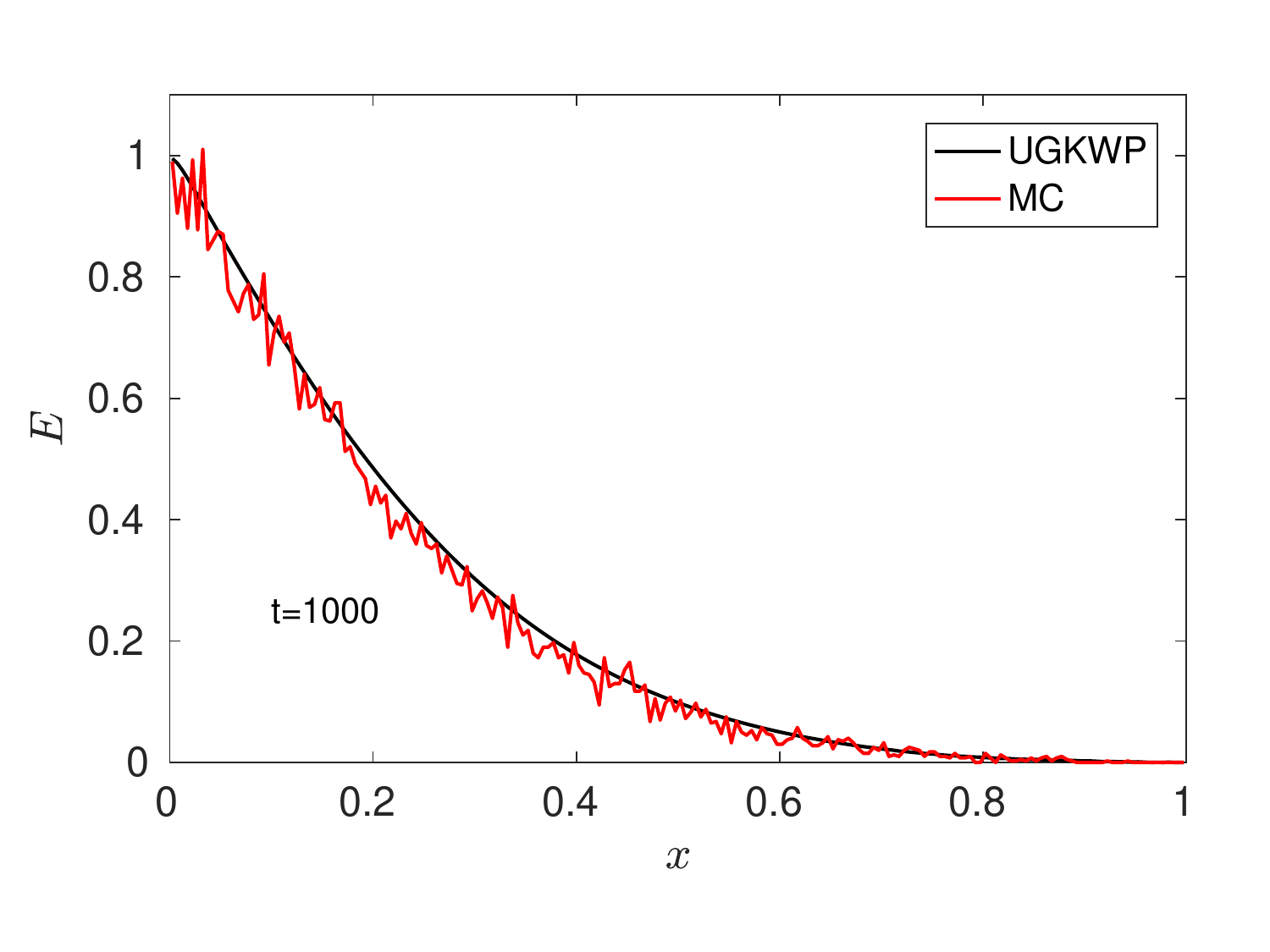}}
  \subfigure[Comparison of $E$ between UGKWP and the Monte Carlo
  solution in thin-to-thick heterogeneous
medium.]{\label{fig:thin2thick-E}
  \includegraphics[width=0.48\textwidth]{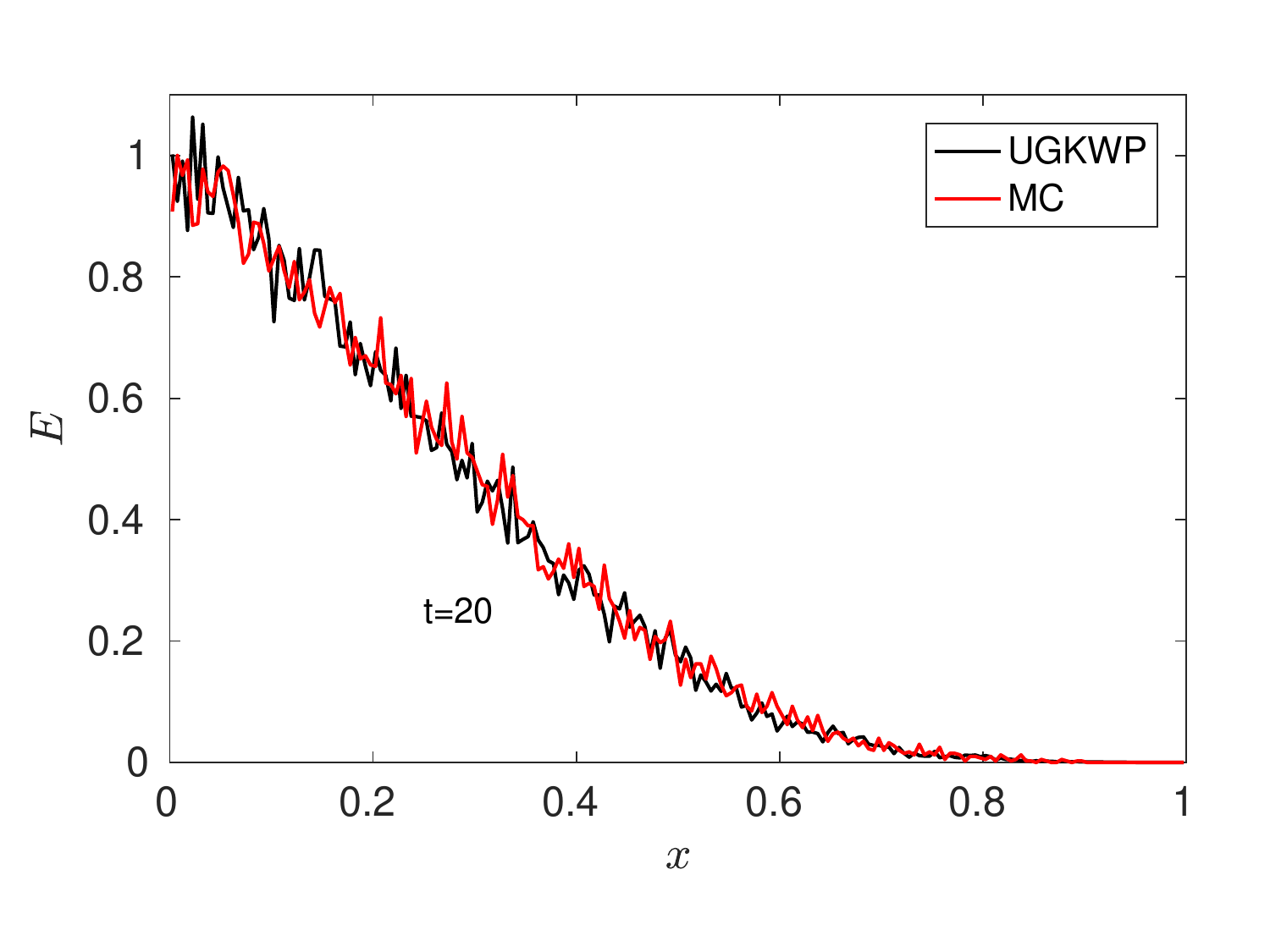}}
  \caption{Numerical results of heterogeneous medium.}
  \label{fig:heterogenous}
\end{figure}

We calculate two test cases. For the first test case, we take $\sigma
= 10000 \text{arctan}(1 - x)$. In this case, radiation passes through
an optically thick  medium near the left boundary and gets into
the region with a gradually reducing of optically thickness. In Figure
\ref{fig:thick2thin-E} we compare the solutions of the UGKWP method and the Monte
Carlo method for $t = 1000$ and they agree with each other closely.
Also, in this case where the medium is relatively optically thick
overall, the UGKWP produces solutions much smoother than those given
by the Monte
Carlo method. For this test
case, the Monte Carlo method uses a maximum of $17972$ particles and
takes $5127$ seconds.  The UGKWP method uses a maxium of $85$
particles and takes $18$ seconds. Therefore, the UGKWP method is much
more efficient than the Monte Carlo method for this test case.

For the second test case, we take 
$\sigma = 100 / \text{arcsin}(1 - x)$, implying that the medium gets more
and more optically thick from the left to right boundaries. In Figure
\ref{fig:thin2thick-E} the numerical result of the UGKWP method is
compared with the solution of the Monte Carlo method at $t = 20$ and
they are almost identical except for the statistical noise. The Monte
Carlo method uses a maximum of $24598$ particles and spends $12$
seconds while the UGKWP method uses a maximum of $24823$ particles and
spends $10$ seconds in computing this case. In this test case the
medium is relatively optically thin overall, and the UGKWP method has
similar time cost as the Monte Carlo method.

\subsection{Marshak wave problem}
This section studies the Marshak wave problem where radiation is
coupled with material medium. Consider the
system
\begin{displaymath}
  \left\{\begin{split}
      & \dfrac{1}{c} \pd{I}{t} + \mu \pd{I}{x}
      = \sigma \left(\dfrac{1}{2}
    a c T^4 - I\right), \\
    & C_v \pd{T}{t} = \sigma \left(\int_{-1}^1
    I(\mu) \dd \mu - a c
  T^4\right).
\end{split}\right.
\end{displaymath}
for semi-infinite domain $x \in [0, +\infty)$ with a computational one $x \in [0,0.5]$. We take absorption coefficient to be $\sigma
= \dfrac{30}{T^3}$, the speed of light to be $c = 29.98$, the
parameter $a = 0.01372$ and the specific heat to be $C_v = 0.3$. The
initial material temperature is set to be $10^{-2}$ and initially
material and radiation energy are at equilibrium. A constant isotropic inflow radiation intensity with an
equivalent radiation temperature of $1$ is imposed on the left
boundary.  Both the UGKWP method and the Monte Carlo method use
$200$ grids in space. For this example, the implicit Monte Carlo
method uses a maximum of $45277$ particles, while the UGKWP method
uses a maximum of $42892$ particles. It takes the implicit Monte
Carlo method $66$ seconds to compute until $t = 1.0$ and the UGKWP
method $60$ seconds.

\begin{figure}[htbp]
  \centering
  \subfigure[Comparison of radiation temperature between UGKWP and the Monte Carlo
  solution for Marshak wave problem.]{
    \label{fig:marshak-radiation}
  \includegraphics[width=0.48\textwidth]{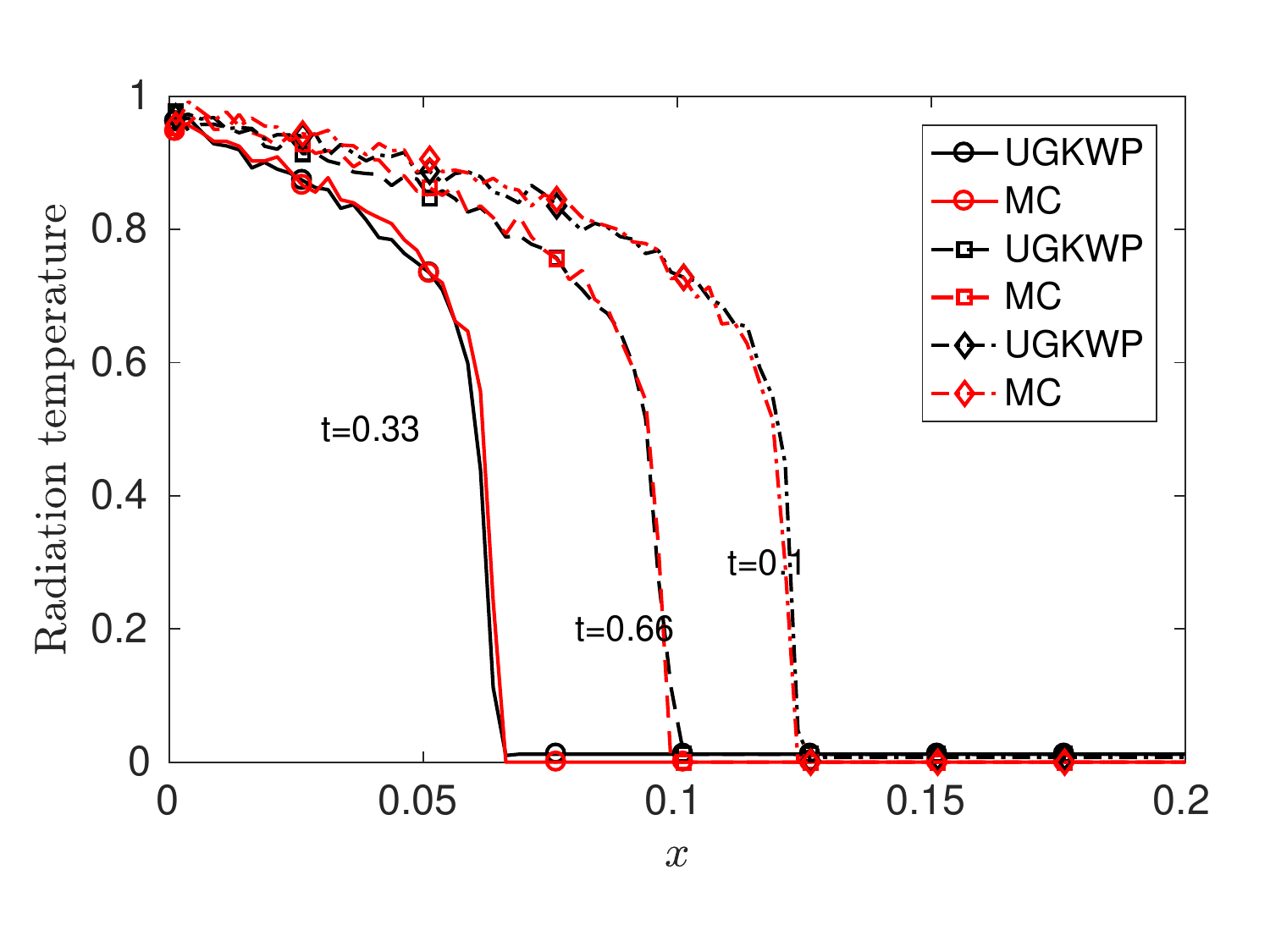}}
  \subfigure[Comparison of material temperature between UGKWP and the Monte Carlo
  solution for Marshak wave problem.]{\label{fig:marshak-material}
  \includegraphics[width=0.48\textwidth]{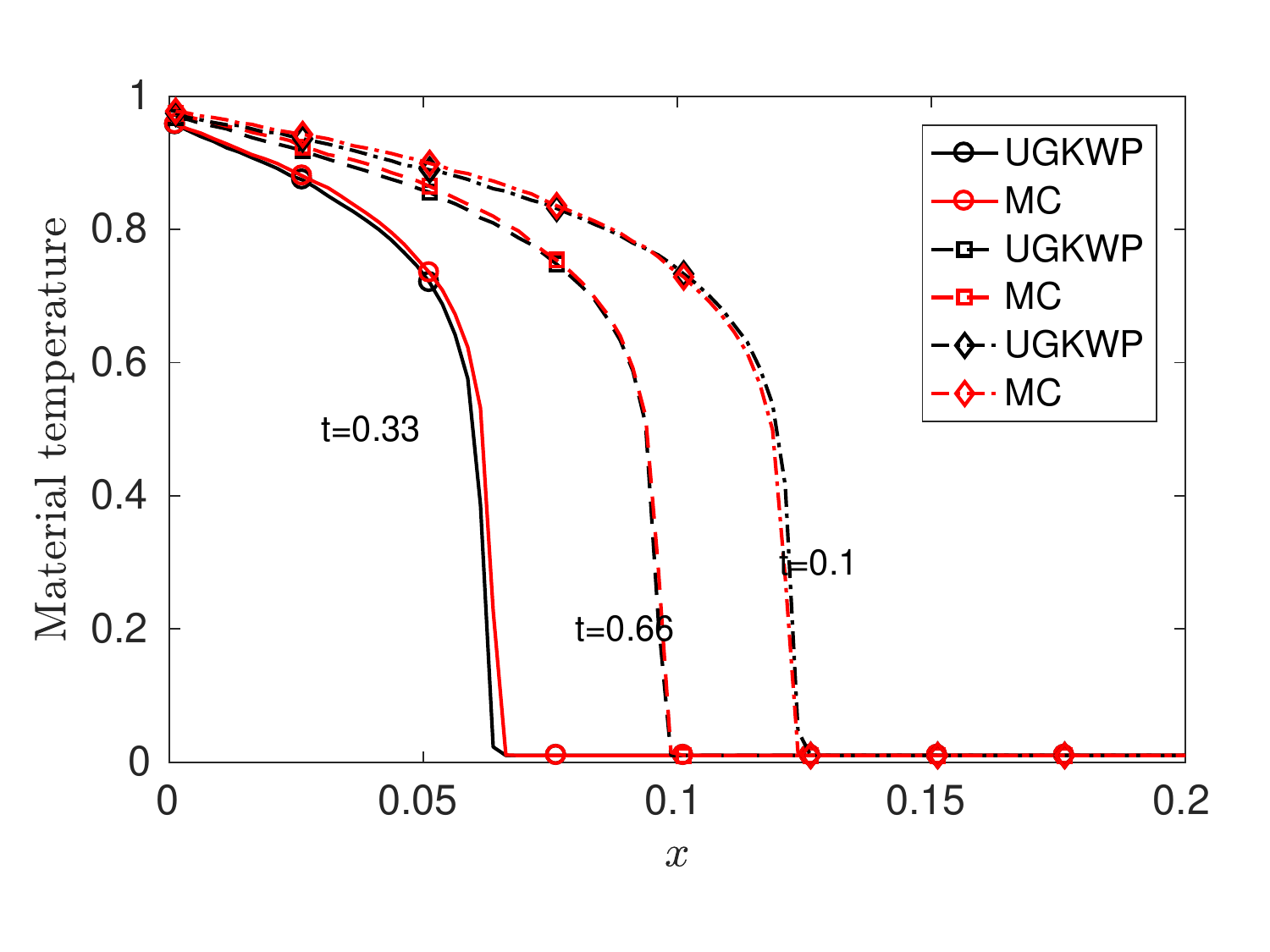}}
  \caption{Numerical results of Marshak wave problem.}
  \label{fig:marshak}
\end{figure}

In Figure \ref{fig:marshak-radiation}, the computed radiation wave front at time
$t = 0.33$, $0.66$ and $1.0$ are given,
while Figure \ref{fig:marshak-material} presents the computed material
temperature. The solutions of the UGKWP method are shown to be consistent with
those of the Monte Carlo method for both radiation and material
temperature.

\subsection{Line-source problem in purely scattering homogeneous
medium}
The next problem we look at is the line-source problem in purely
scattering medium. In this test case, we consider equation \eqref{eq:rt-nondimensional}
for $\epsilon = 1$ and $c = 1$.
The spatial 2D computational domain is $[-1.5,1.5] \times [-1.5, 1.5]$. The initial density
distribution is
\begin{displaymath}
  I(0, \bx, \bsOmega) = \dfrac{1}{4\pi} \delta(x) \delta(y),
\end{displaymath}
which means that initially all particles are concentrated at $x = y =0$ and they spread out over the time.
This is a particularly difficult problem for the $S_n$ based methods
because they suffer from severe ray effect. Previous studies have used
this problem for comparisons between methods \cite{brunner2002forms} and as a test case for
schemes to mitigate the ray effect \cite{camminady2019ray}.

Both the UGKWP method and the Monte Carlo method use $201 \times 201$
cells in space. The UGKWP method uses $1.633 \times 10^7$ particles
while the Monte Carlo method uses $1.638\times 10^7$ particles and the
same spatial mesh as the UGKWP method. Because this example considers
the kinetic regime, the number of particles used in the UGKWP method
is almost the same as that used in the Monte Carlo method and the
computation time is similar, with $1530$ seconds for the UGKWP method
and $1034$ seconds for the Monte Carlo method. The UGKWP method is
slightly more expensive due to its additional overheads.

\begin{figure}[htbp]
  \centering
  \subfigure[Contour plot of the solution of $E$ by the Monte Carlo method
  as a function of the spatial coordinate in the line-source problem at $t = 1.2$.]{
    \label{fig:contour-linesource-mc}
  \includegraphics[width=0.48\textwidth]{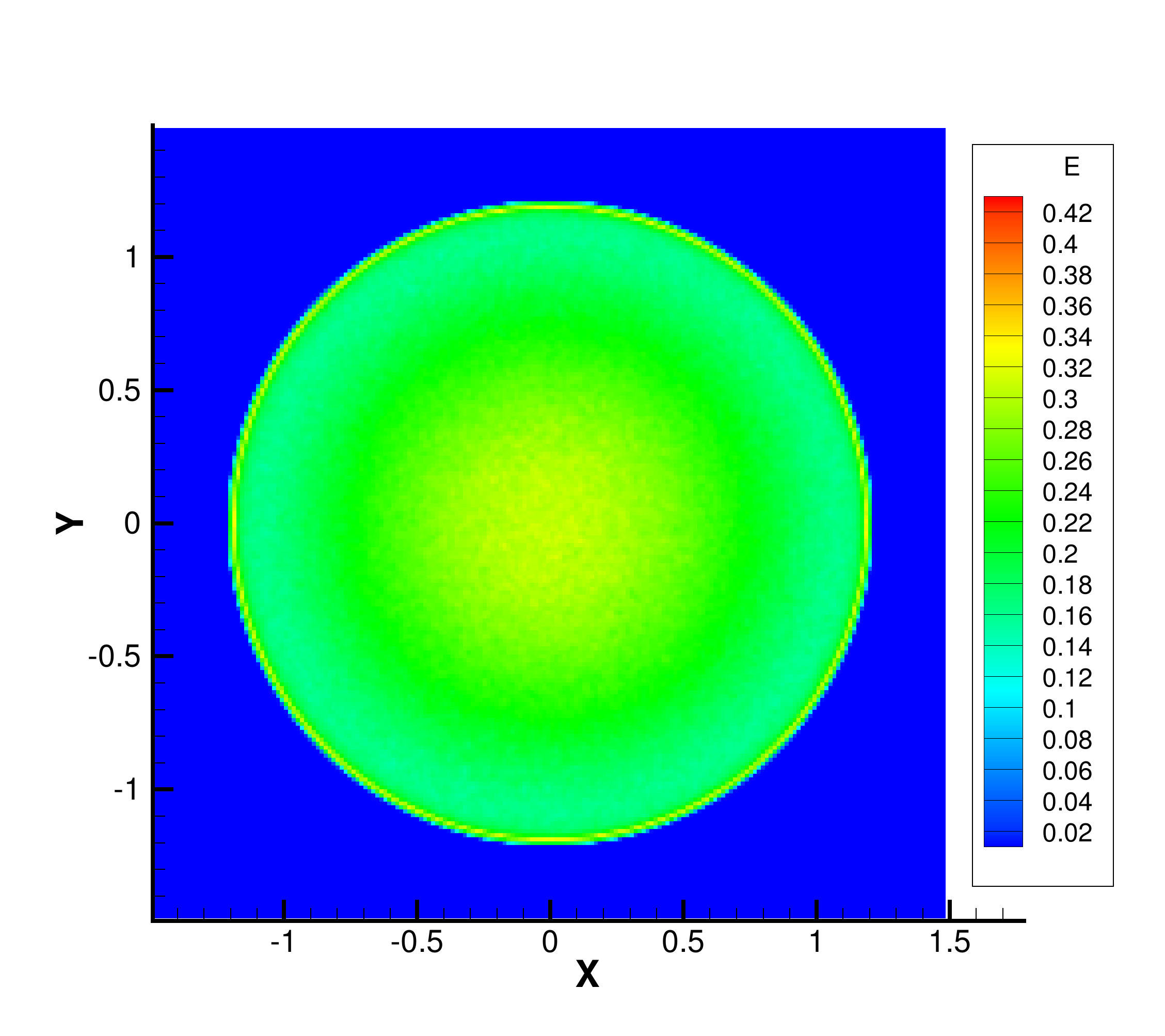}}
  \hfill
  \subfigure[Contour plot of the solution of $E$ by the UGKWP method
  as a function of the spatial coordinate in the line-source problem at $t = 1.2$.]{
    \label{fig:contour-linesource-ugkwp}
  \includegraphics[width=0.48\textwidth]{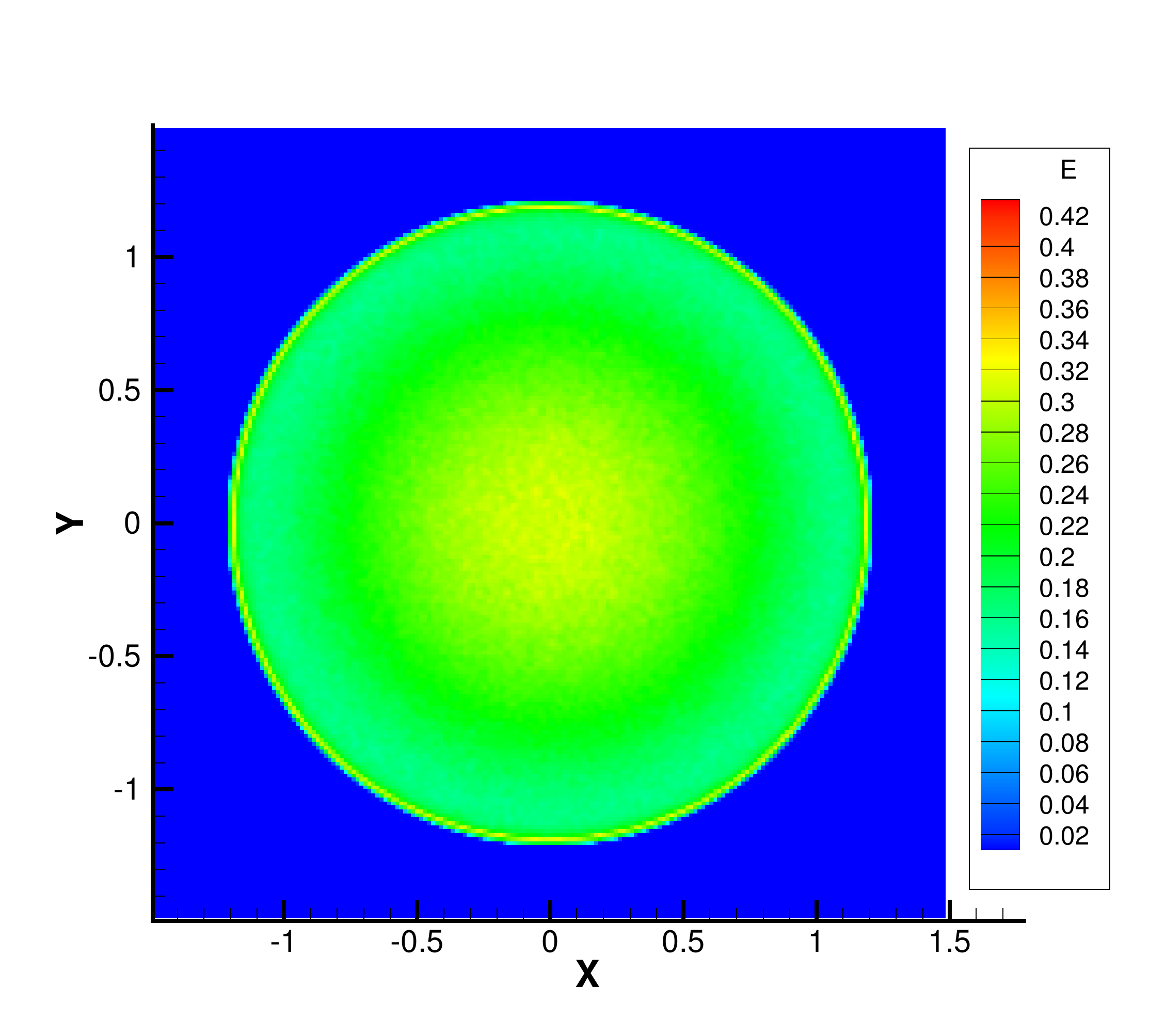}}
 \hfill
  \subfigure[Contour plot of the solution of $E$ by the $S_8$ method
  as a function of the spatial coordinate in the line-source problem at $t = 1.2$.]{
    \label{fig:contour-linesource-s8}
  \includegraphics[width=0.48\textwidth]{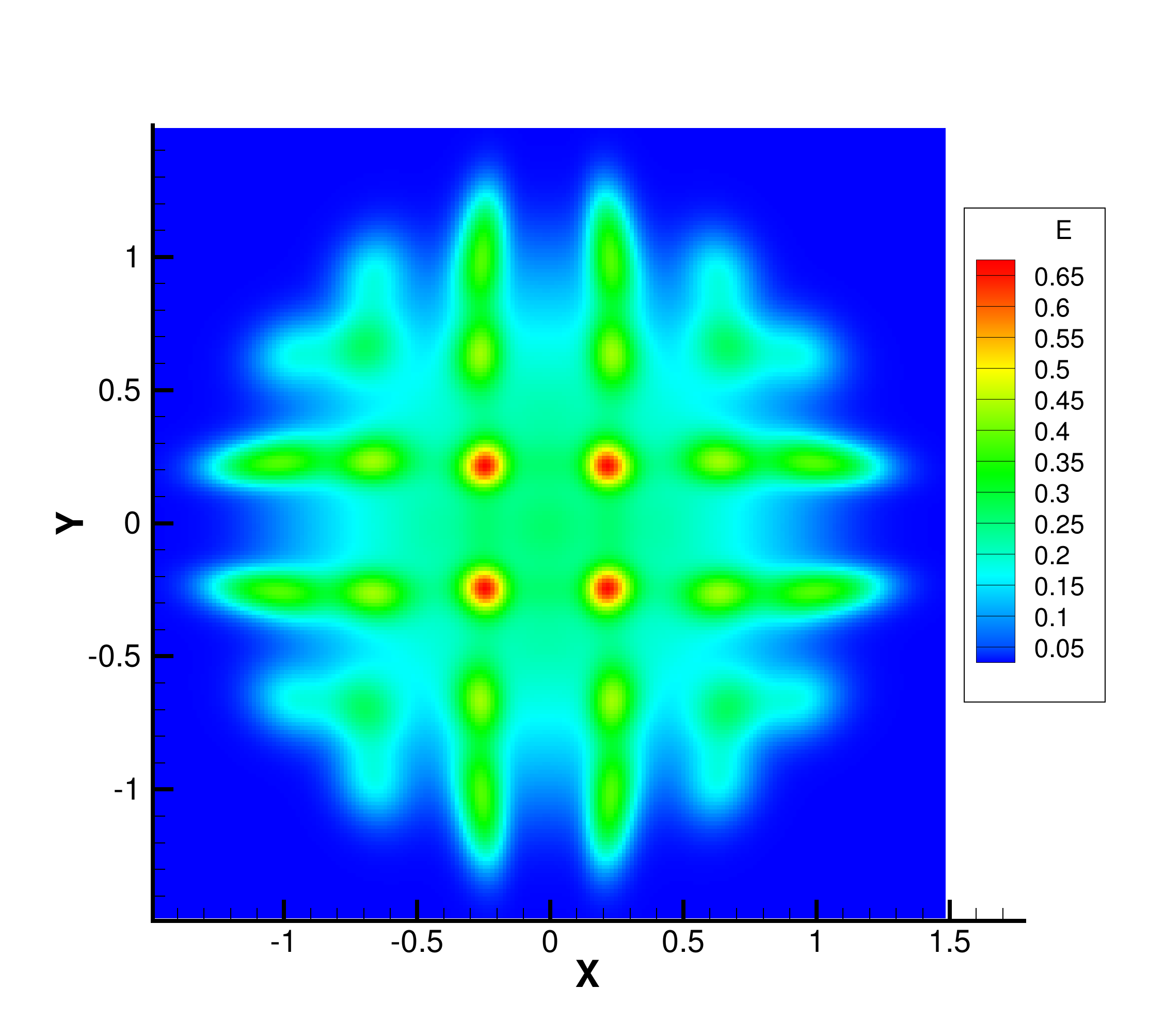}}
  \hfill
  \subfigure[Slice at $y=0$ of the solution of $E$ by the Monte Carlo
    method and the UGKWP method as a function of $x$ in the
  line-source problem at $t = 1.2$.]{
    \label{fig:slice-linesource-compare}
  \includegraphics[width=0.48\textwidth]{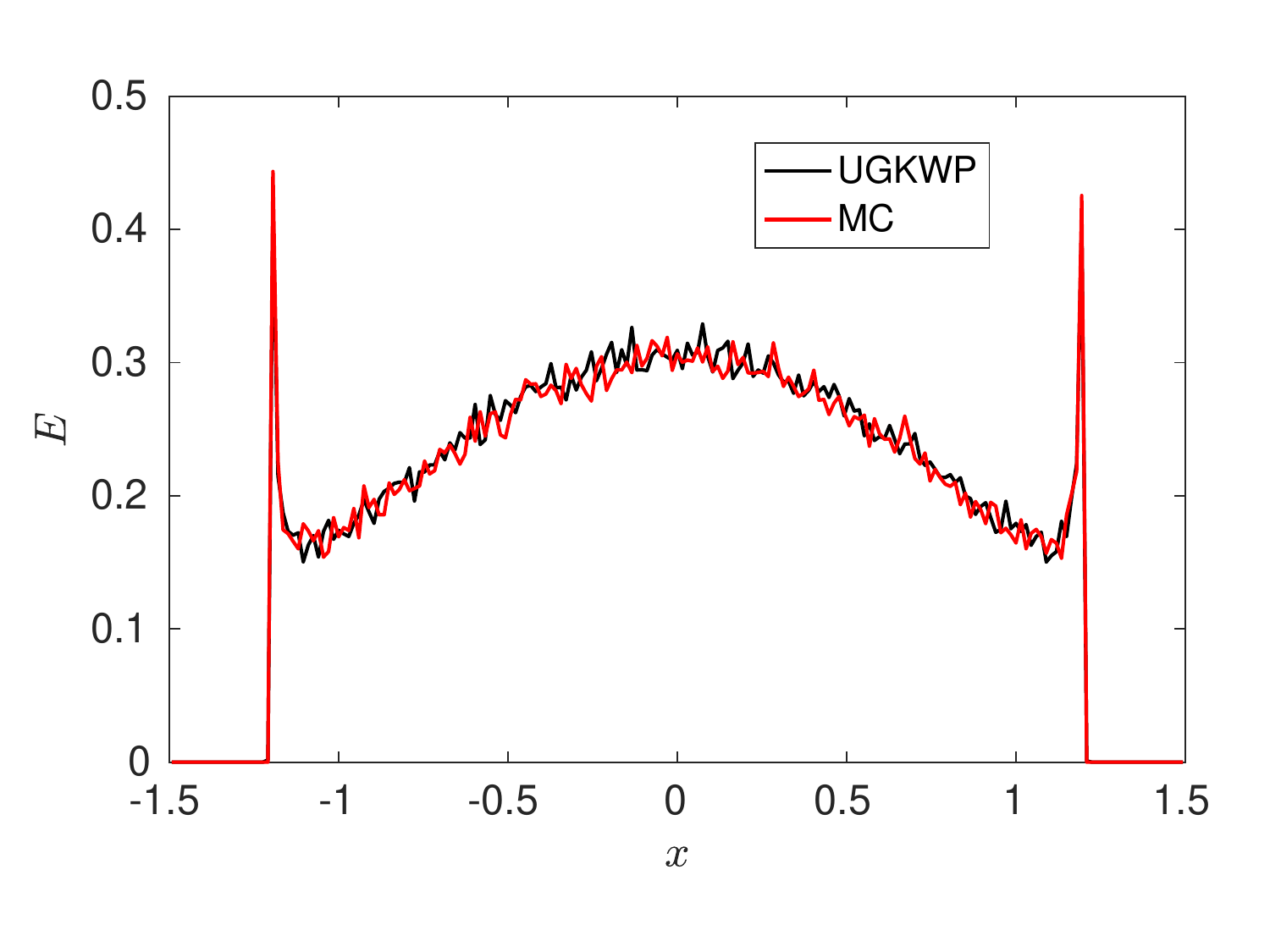}}
  \caption{Numerical results of the line-source problem.}
  \label{fig:linesource}
\end{figure}

Fig.\ref{fig:linesource} compares the solution of $E$ between the UGKWP
method, the Monte Carlo method and the $S_8$ method for time $t =
1.2$. For the $S_8$ computation, we use an equal-weight quadrature
set, $201\times 201$ cells for the spatial mesh and the DOM-UGKS. 
The UGKWP solution is consistent with
the Monte Carlo solution. Both solutions display a sharp spike near
the wave front and has a smooth nonzero region behind the wave front,
and neither suffers from the ray effect. On the other hand, the $S_8$
solution is qualitatively incorrect due to the ray effect. This
example demonstrates that the UGKWP method preserves the advantage of
the Monte Carlo method in the rarefied regime without suffering from the ray effect.

\subsection{Emitting isotropic source in lattice medium}
\begin{figure}[htbp]
  \centering
  \includegraphics[width=0.48\textwidth]{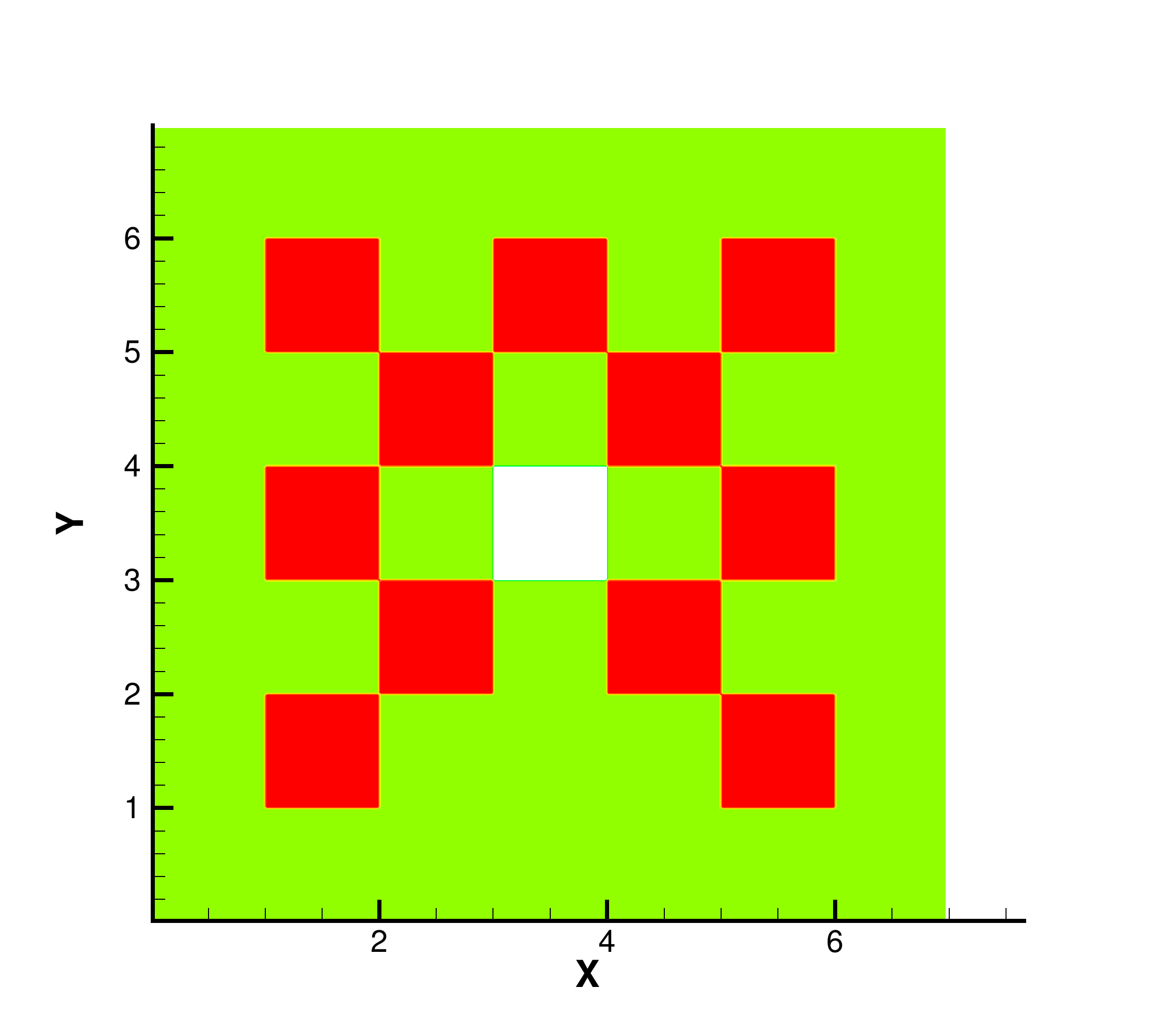}
  \caption{Layout of the lattice problem.}
  \label{fig:lattice_mesh}
\end{figure}

\begin{figure}[htbp]
  \centering
  \subfigure[Contour plot of the solution of $\ln E$ by the Monte Carlo method
  as a function of the spatial coordinate in the lattice problem at $t = 2$.]{
    \label{fig:contour-lattice-orig-mc}
  \includegraphics[width=0.48\textwidth]{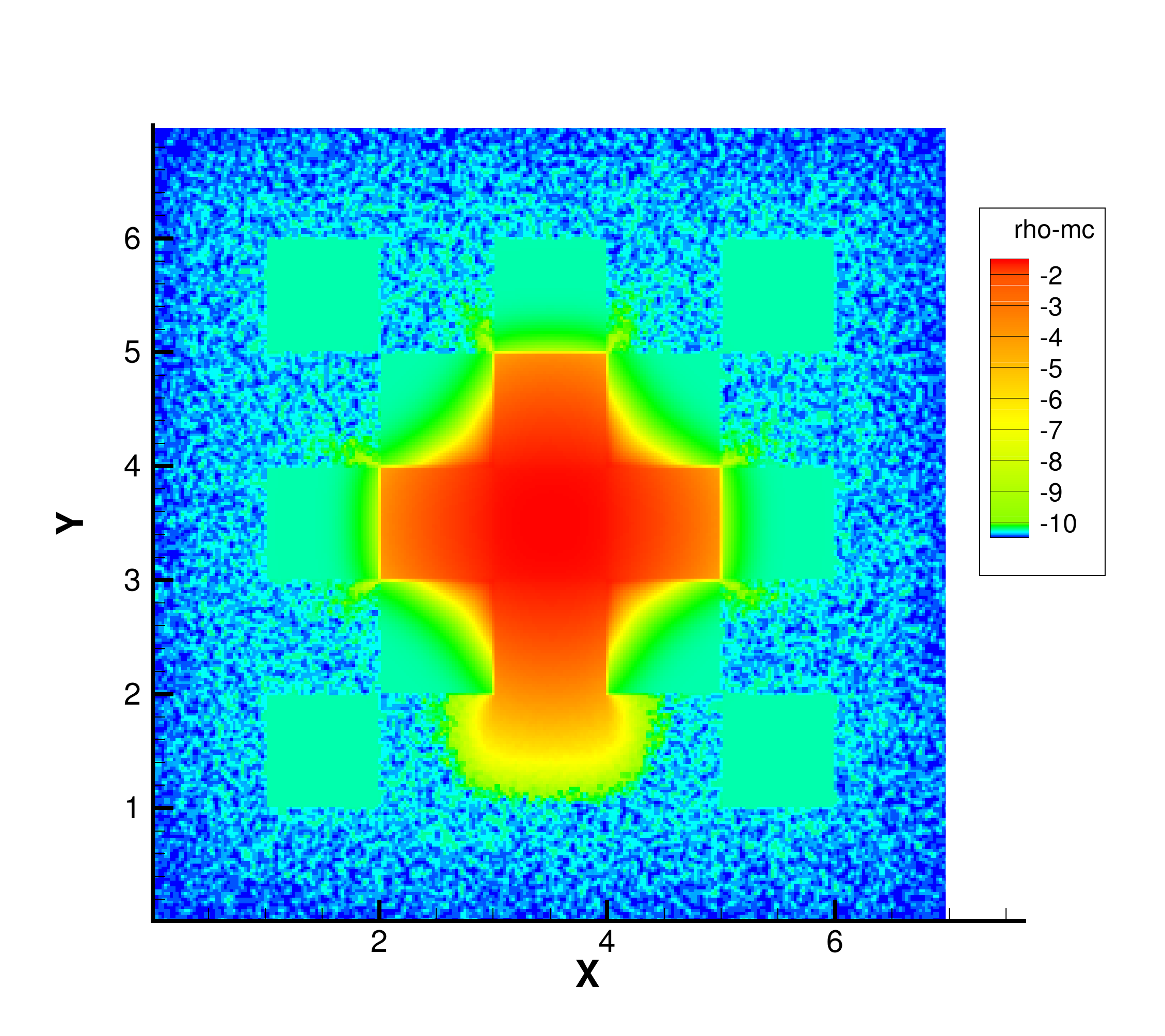}}
 \hfill
 \subfigure[Contour plot of the solution of $\ln E$ by the UGKWP method
  as a function of the spatial coordinate in the lattice problem at $t = 2$.]{
    \label{fig:contour-lattice-orig-ugkwp}
  \includegraphics[width=0.48\textwidth]{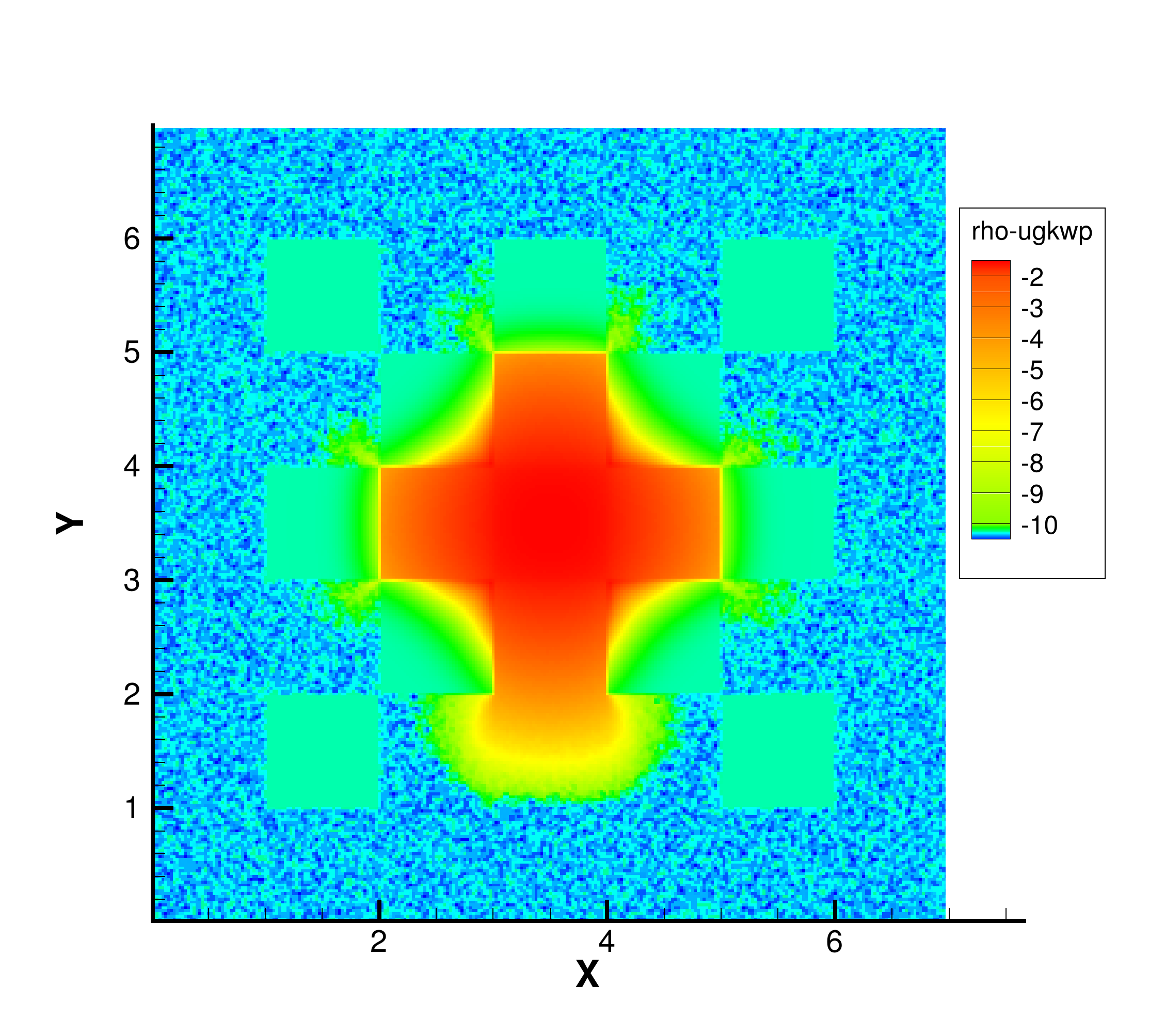}}
  \subfigure[Slice at $y = 0$ of the solution of $\ln E$ by the Monte
    Carlo method and UGKWP as a function of $x$ in the lattice problem
  at $t = 2$.]{
    \label{fig:slice-y-0}
  \includegraphics[width=0.48\textwidth]{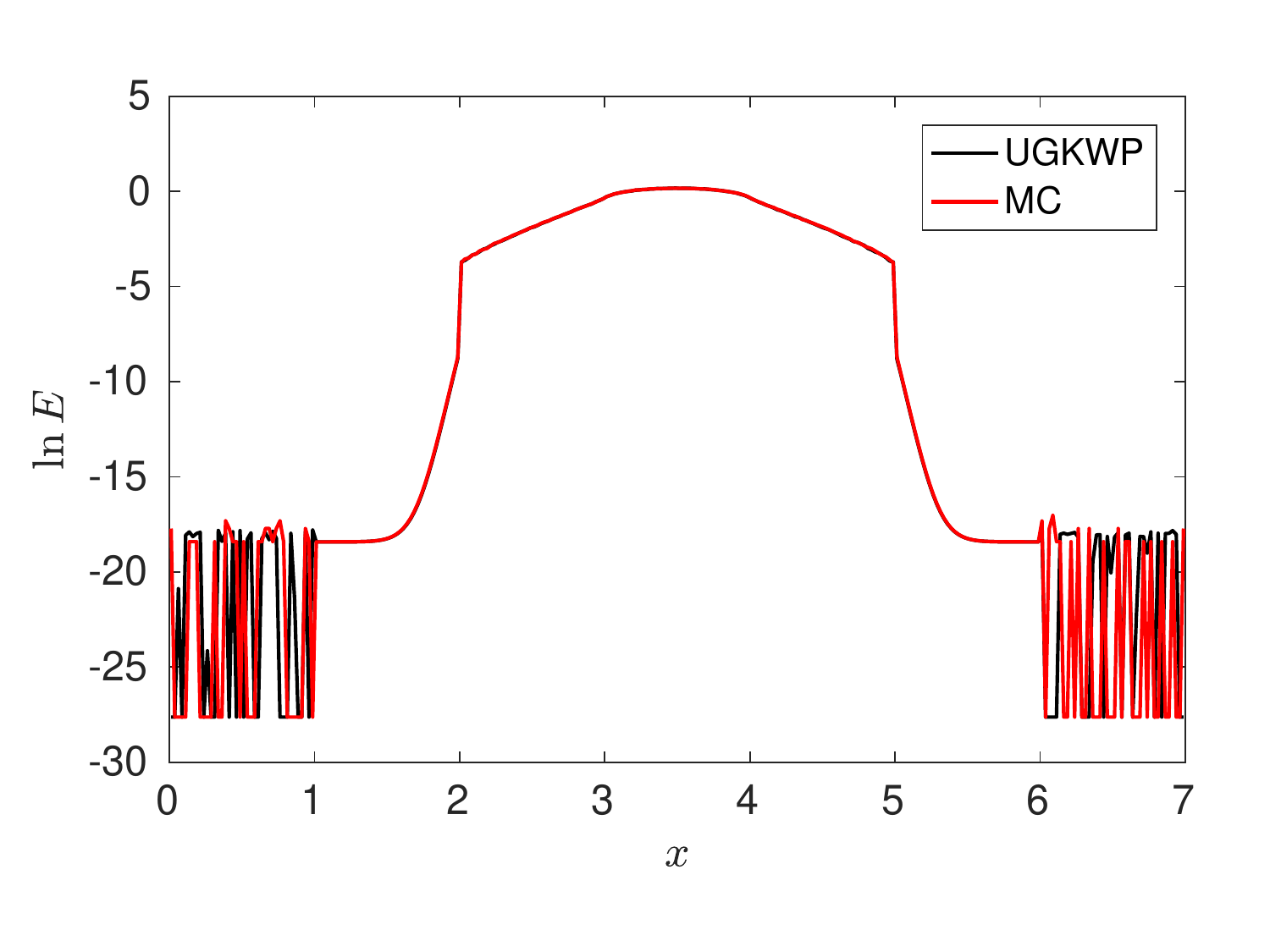}}
 \hfill
 \subfigure[Slice at $x = 0$ of the solution of $\ln E$ by the Monte
    Carlo method and UGKWP as a function of $y$ in the lattice problem
  at $t = 2$.]{
    \label{fig:slice-x-0}
  \includegraphics[width=0.48\textwidth]{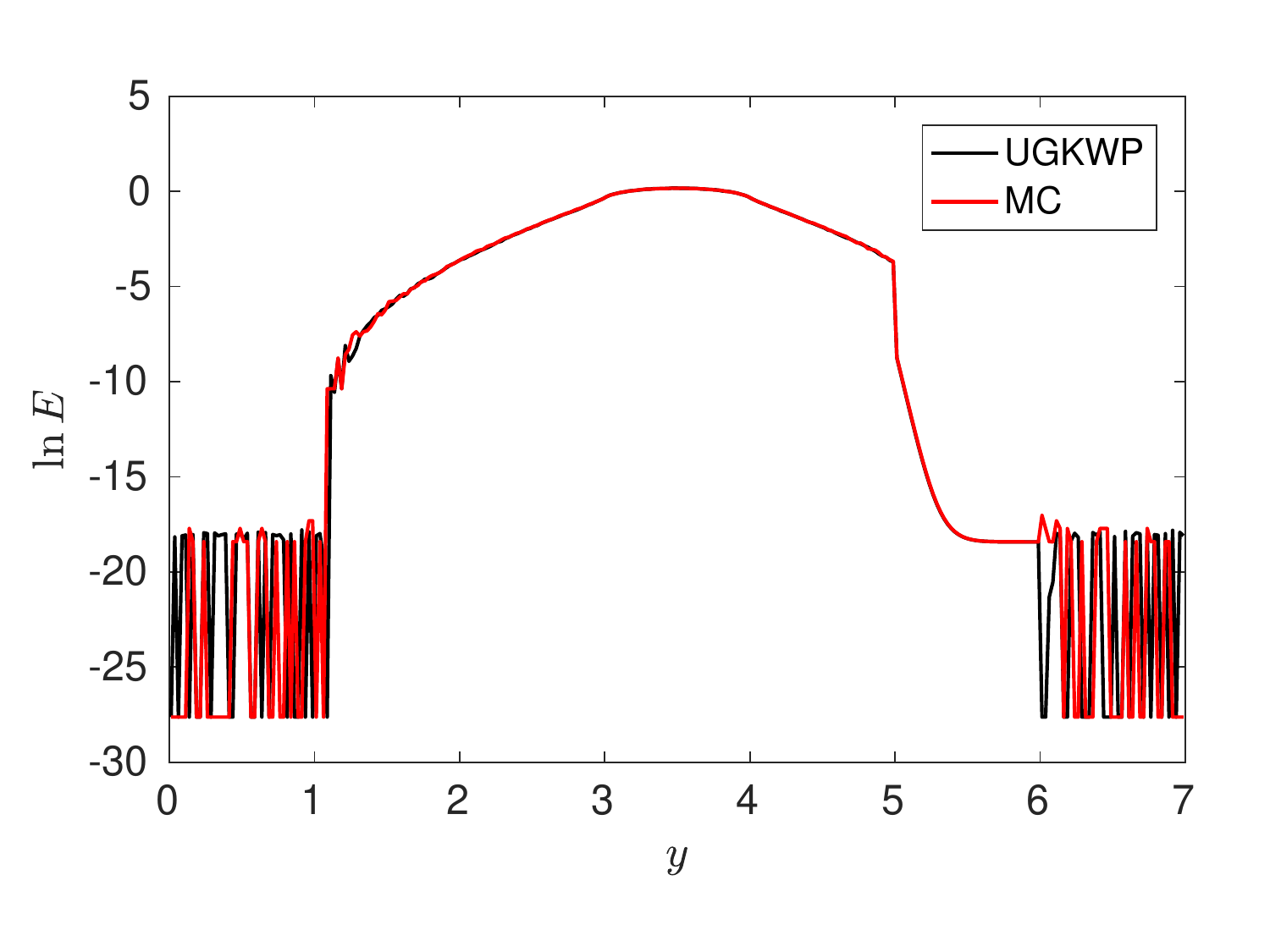}}
 \hfill
 \caption{Numerical results of the lattice problem.}
  \label{fig:lattice-new}
\end{figure}

In this section we study a problem with multiple medium
\cite{mcclarren2010robust} by considering the following equation
\begin{displaymath}
  \left\{\begin{split}
  & \dfrac{\epsilon^2}{c} \pd{I}{t} + \epsilon\bsOmega \cdot \nabla I
      = \sigma_a \left(\dfrac{1}{4\pi}
      a c T^4 - I\right) + \sigma_s \left(\dfrac{1}{4\pi} E - I\right)
      + \epsilon^2 G, \\
    & \epsilon^2 C_v \pd{T}{t} = b \Delta T + \sigma_a \left(\int_{\bbS^2}
    I(\bsOmega) \dd\bsOmega - a c
  T^4\right).
\end{split}\right.
\end{displaymath}
The diffusion term $b \Delta T$ is solved
by using a backward Euler method.
The two-dimensional physical space $[0, 7]
\times [0, 7]$ consists of a set of squares belonging to a strongly
absorbing medium and a background of weakly scattering medium. The
specific layout of the problem is given in Fig.\ref{fig:lattice_mesh}, where the light green regions and the white
region are purely scattering with $\sigma_s = 2$ and $\sigma_a = 0$;
the red regions are pure absorbers with $\sigma_s = 0$ and $\sigma_a =
2000$. In the white region in the center there is an isotropic source
$G = \dfrac{1}{4\pi}$. The initial material temperature is $10^{-2}$ and initially the
radiation and material temperature are at equilibrium.
We take
$C_v = 0.3$, $b = 0.03$ and $c = a = \epsilon = 1$ to simulate the
case.
Both the UGKWP method and the Monte Carlo method use a mesh of $280
\times 280$ in physical space. The results at time $t = 2$ are
summarized in Fig. \ref{fig:lattice-new} with the natural logarithm
of density distribution shown in contour and slice, and they are in
agreement with each other. This is a test case which includes material
in both the kinetic and the diffusive regimes, but most of the time
particles travel in the kinetic regime.  The Monte Carlo method uses
$10^8$ particles and took $5692$ seconds, while the UGKWP method uses
$8.5\times 10^7$ particles and took $3092$ seconds, showing the UGKWP
method to be more efficient than the Monte Carlo method for this case
where there is mixed medium.

\subsection{A hohlraum problem}
\begin{figure}[htbp]
  \centering
  \includegraphics[width=0.48\textwidth]{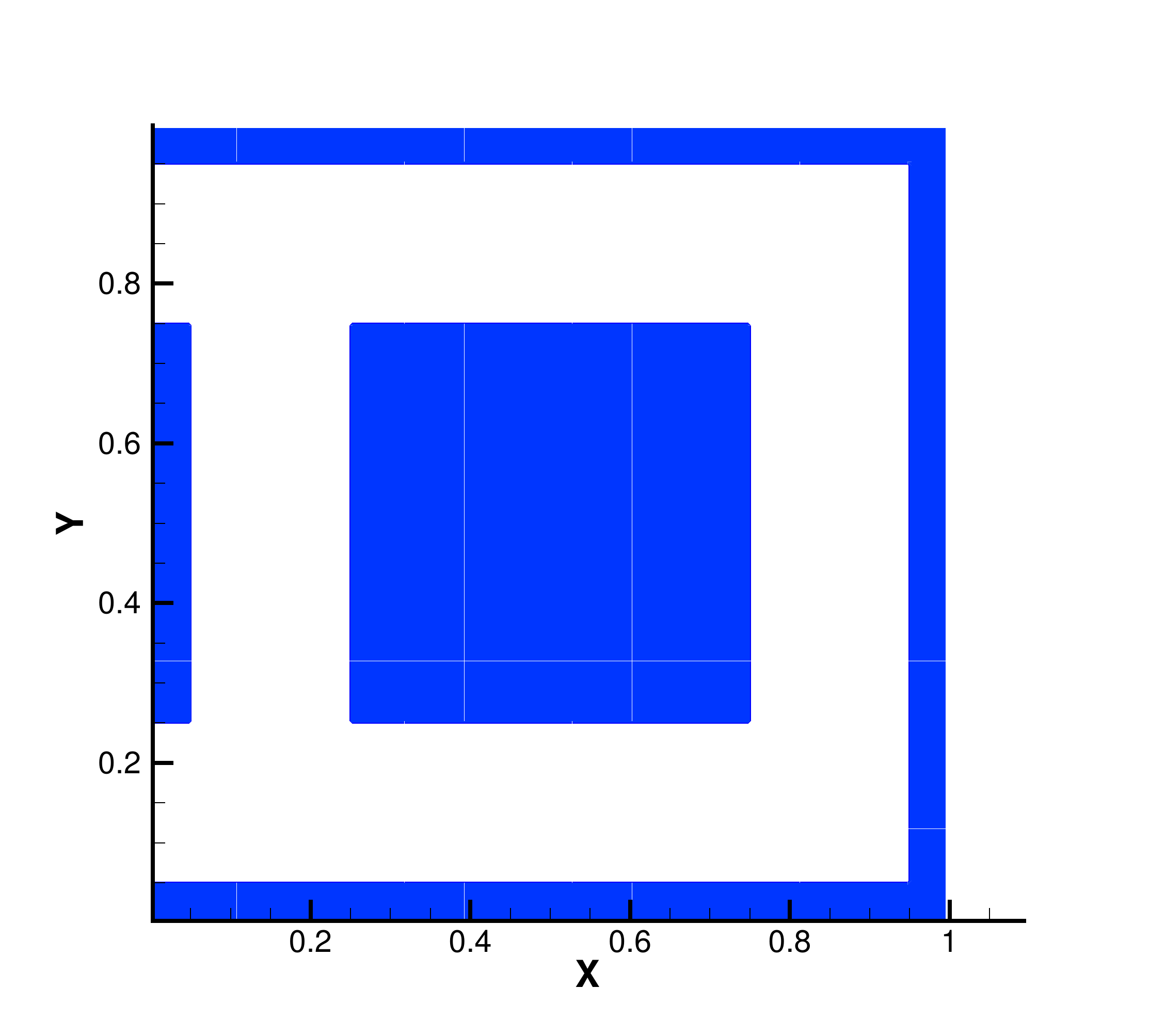}
  \caption{Layout of the hohlraum problem. In the blue regions
  $\sigma_a = 1000 T^{-3}$, $\sigma_s = 0$ and $C_v = 0.3$. The white
region is vacuum, $\sigma_a = \sigma_s = 0$.}
  \label{fig:hohlraum_coeff}
\end{figure}

This examples considers a hohlraum problem similar to the one studied
in \cite{mcclarren2010robust}. We study equation \eqref{eq:rt-coupled}
with $G = 0$.  The problem layout is given in Fig.\ref{fig:hohlraum_coeff}. This is a problem of purely absorbing medium
where the absorption coefficient relies on the material temperature.
As the material temperature varies, the values of the absorption coefficient cover a
wide range, presenting challenges to numerical methods.
On the left boundary there is an isotropic source with the specific
intensity $I(\bsOmega) = \dfrac{1}{4\pi}$. Initially radiation and
material temperature are at equilibrium and the initial material
temperature is $T = 10^{-2}$.

The UGKWP method uses $400 \times 400$ grids in space while the Monte
Carlo method use $800 \times 800$
grid in space. We compute for $t = 2$. The Monte Carlo method uses a
maximum of $2 \times 10^7$ particles and takes $9.8$ hours. The
UGKWP method uses a maximum of $3 \times 10^6$ particles and takes
$1406$ seconds, therefore the UGKWP method is $25$ times more efficient than the
Monte Carlo method for this test case. For this test case, the initial
absorption coefficient on the boundary is extremely large, and the Monte
Carlo method took an especially long time over the first time step when the
inflow radiation first heats up the boundary material. The UGKWP
method, on the other hand, has no such problems.

\begin{figure}[htbp]
  \centering
 \subfigure[Contour plot of the solution of $E$ by the Monte Carlo method
  as a function of the spatial coordinate in the hohlraum problem
at $t = 2$.]{
    \label{fig:contour-hohlraum-E-mc-new}
  \includegraphics[width=0.48\textwidth]{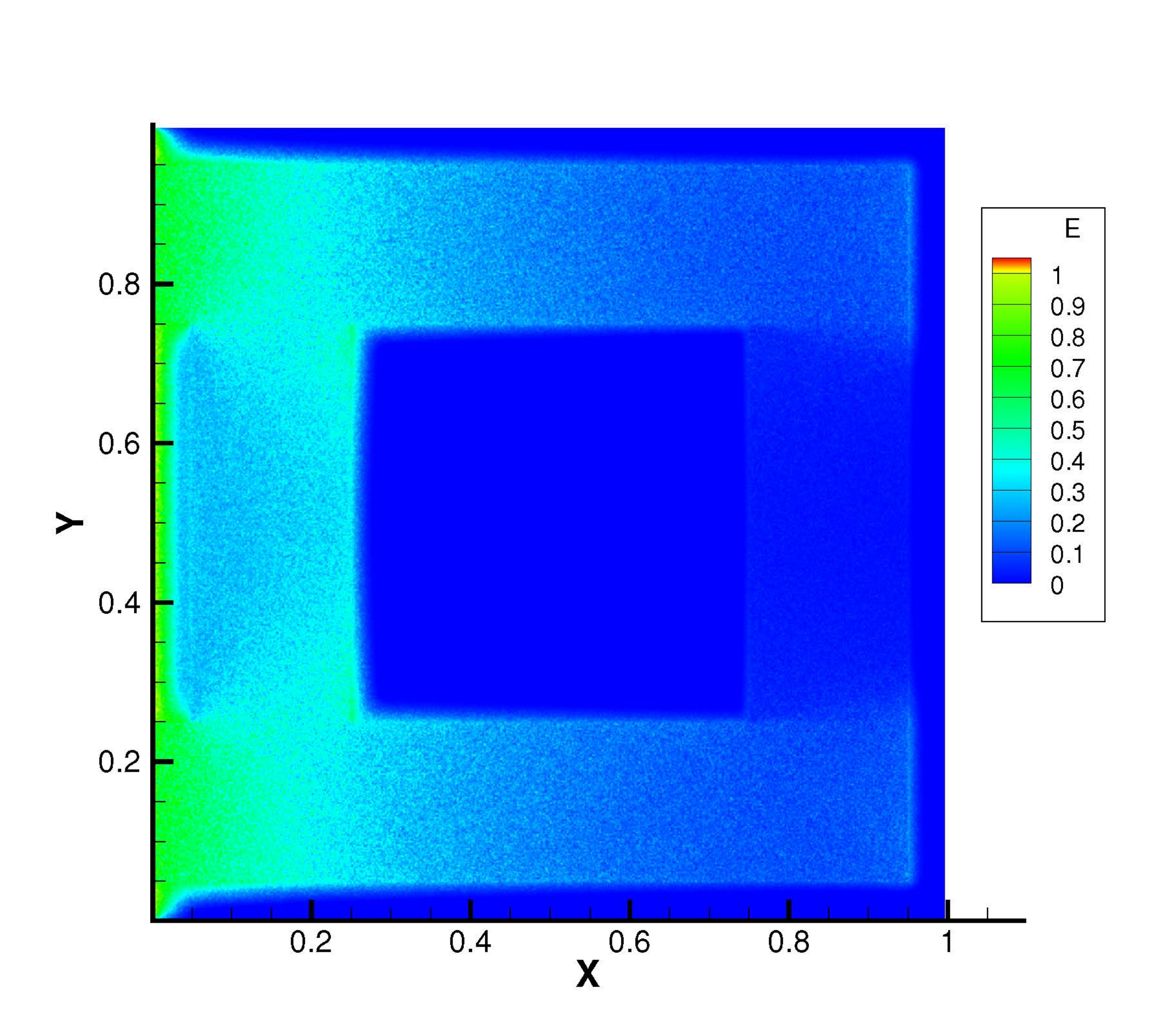}}
 \hfill
 \subfigure[Contour plot of the solution of $E$ by the UGKWP method
  as a function of the spatial coordinate in the hohlraum problem with
  at $t = 2$.]{
    \label{fig:contour-hohlraum-E-ugkwp-new}
  \includegraphics[width=0.48\textwidth]{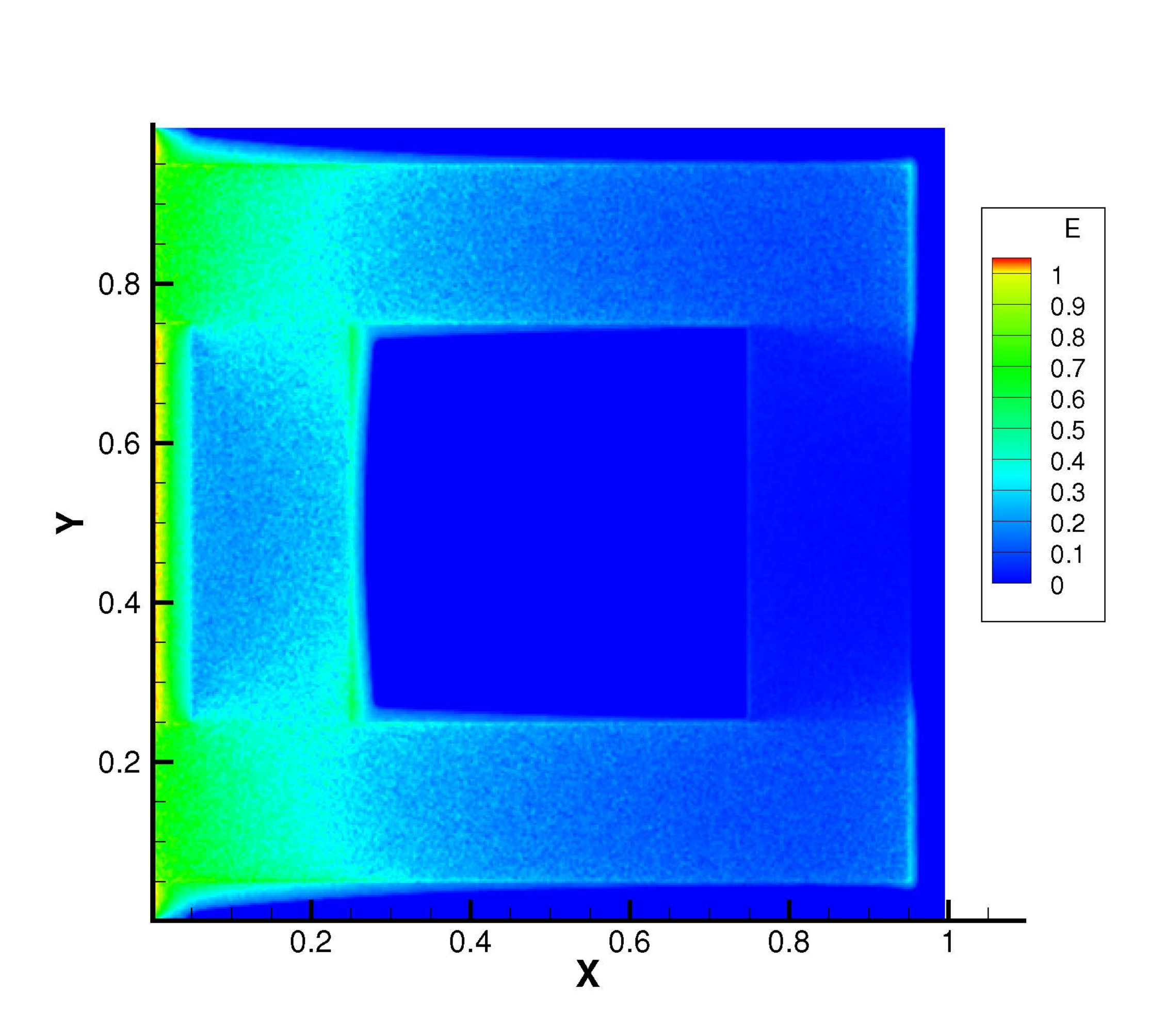}}
 \subfigure[Contour plot of the solution of $T$ by the Monte Carlo method
  as a function of the spatial coordinate in the hohlraum problem
at $t = 2$.]{
    \label{fig:contour-hohlraum-T-mc-new}
  \includegraphics[width=0.48\textwidth]{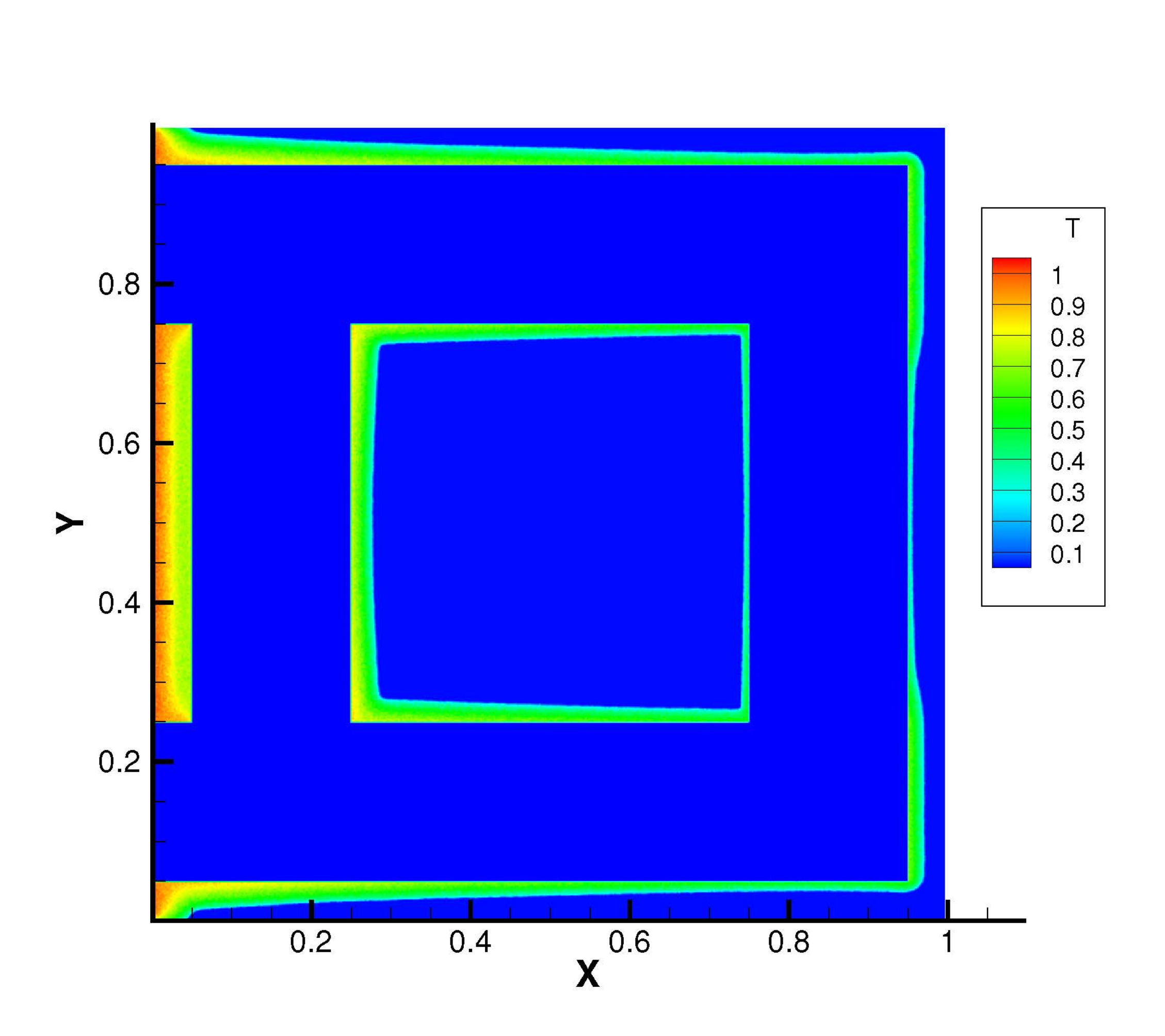}}
 \hfill
 \subfigure[Contour plot of the solution of $T$ by the UGKWP method
  as a function of the spatial coordinate in the hohlraum problem with
  at $t = 2$.]{
    \label{fig:contour-hohlraum-T-ugkwp-new}
  \includegraphics[width=0.48\textwidth]{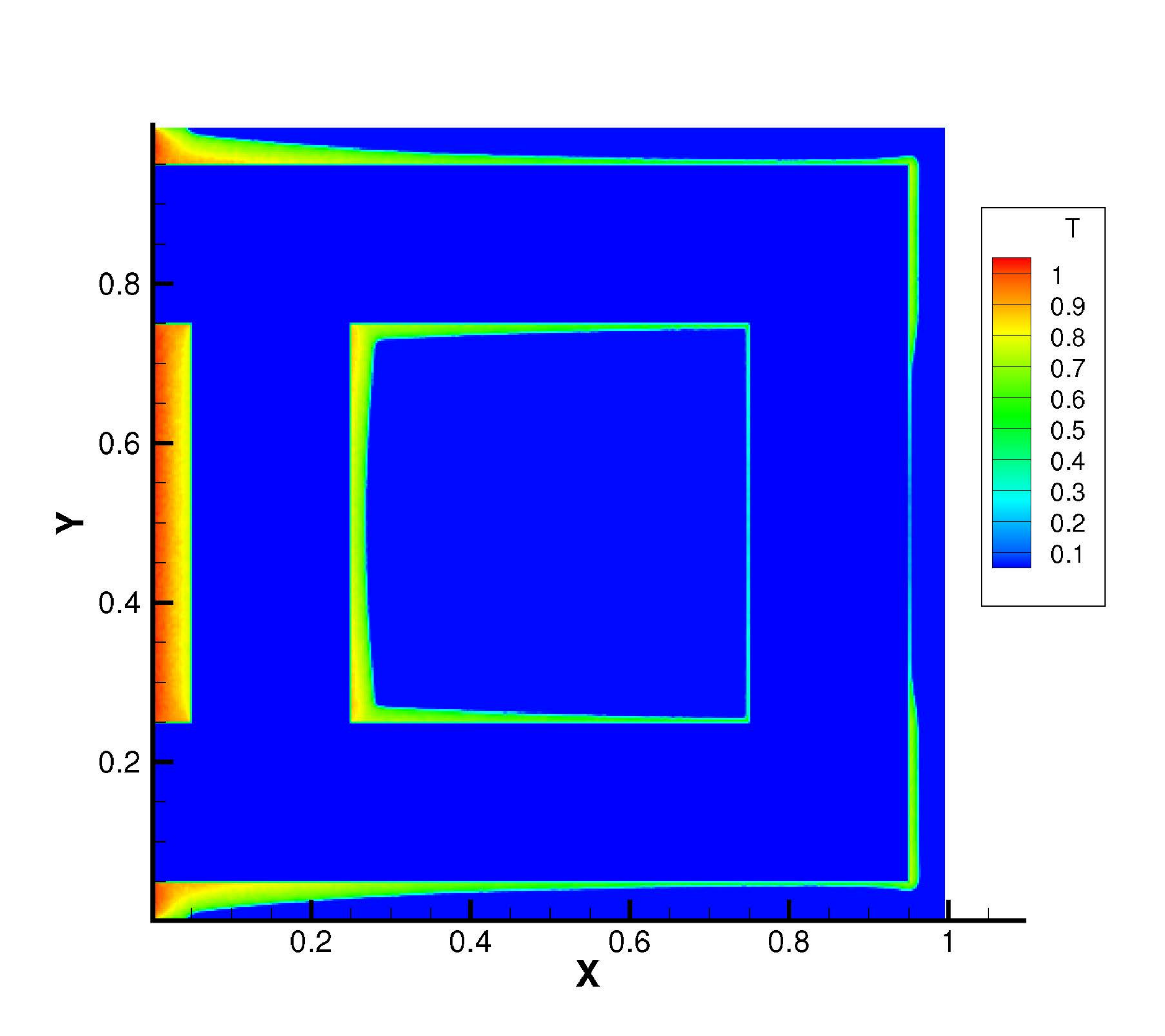}}
  \caption{Numerical results of the hohlraum problem I.}
  \label{fig:hohlraum1}
\end{figure}

\begin{figure}[htbp]
  \centering
   \subfigure[Comparison of sliced solution at $y=0$ of the solution of
   $E$ by the UGKWP method and the Monte Carlo method
  as a function of $x$ coordinate in the hohlraum problem
at $t = 2$.]{
    \label{fig:slice-hohlraum-E-1000}
  \includegraphics[width=0.48\textwidth]{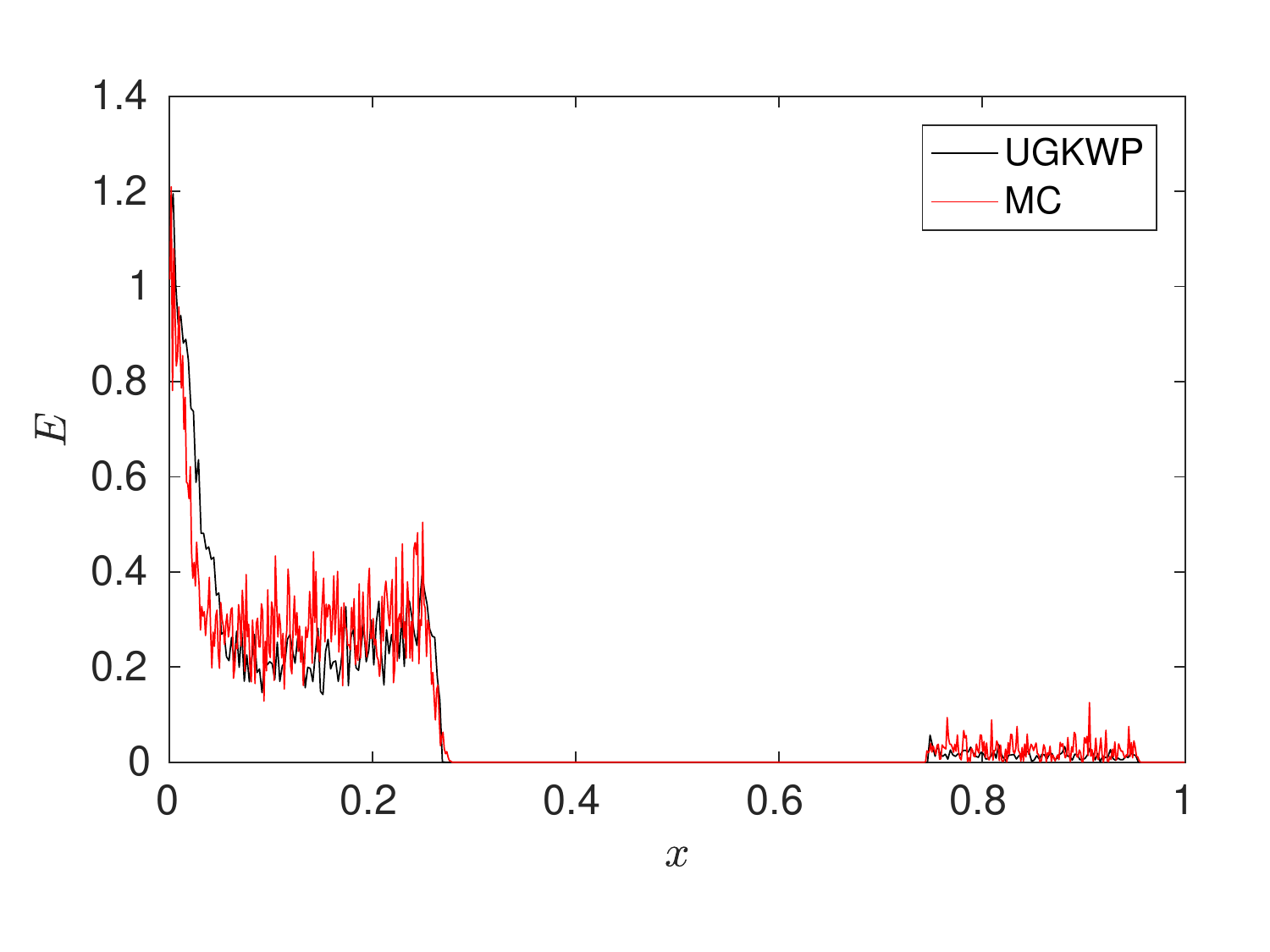}}
 \hfill
 \subfigure[Comparison of sliced solution at $y=0$ of the solution of
   $T$ by the UGKWP method and the Monte Carlo method
  as a function of $x$ coordinate in the hohlraum problem at $t = 2$.]{
    \label{fig:slice-hohlraum-T-1000}
  \includegraphics[width=0.48\textwidth]{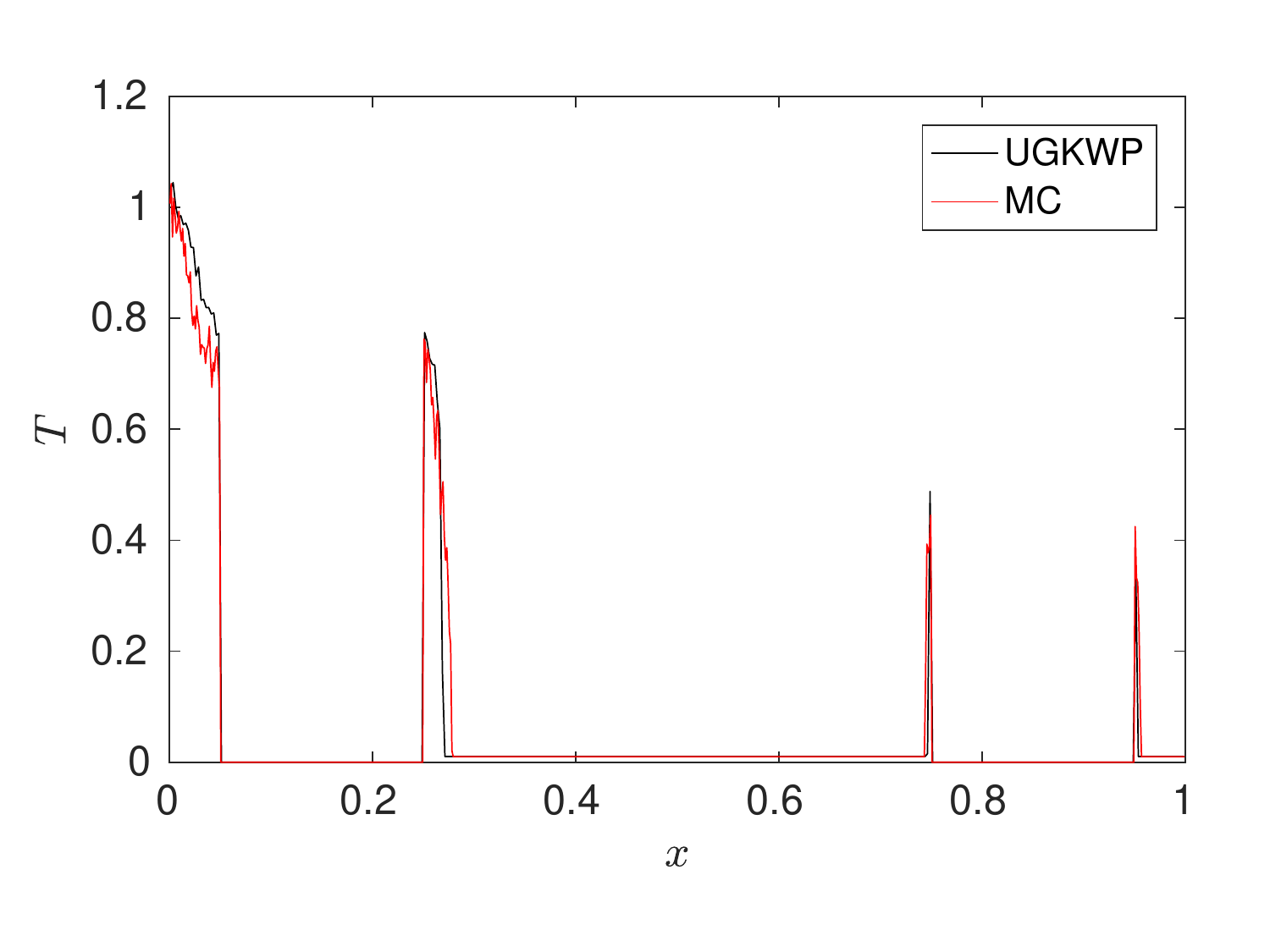}}
  \caption{Numerical results of the hohlraum problem II.}
  \label{fig:hohlraum2}
\end{figure}

The authors of \cite{mcclarren2010robust} pointed out solution to this
problem should contain two characteristics: the central block should
be heated non-uniformly; and the region behind the block with respect
to the source should have less radiation energy then regions within
the line of sight of the source. Their studies showed that the
diffusion equation could not capture these two characteristics. In
Fig. \ref{fig:hohlraum1} and \ref{fig:hohlraum2}, we compare
the solution of the UGKWP method and the Monte Carlo method for energy
density and material temperature at $t = 2$ and they are all
consistent with each other. Both solutions agree with each other such
that there is less radiation to the right of the block. This shows
that the UGKWP method keeps the accuracy of the Monte Carlo method and
is more accurate than the diffusion approximation in problems which
are essentially multiscale in nature.

\section{Conclusion}\label{sec:conclude}
In this paper, for the first time a unified gas-kinetic wave-particle
method is proposed to simulate photon transport. 
Due to the particle wave decomposition and their dynamic coupling, the UGKWP method becomes 
a multiscale method to simulate transport physics with the most efficient way in different regimes. 
For the linear transport equation, this method recovers the solution
of the diffusion equation in the optically thick limit without
constraint on the time step being less than the photon's mean
collision time. At the same time, it gives the exact solution in the
free transport regime.
In the transition regime, both macroscopic wave and kinetic particle contribute to the radiative transfer, and their weights and
coupling depend on the ratio of the time step to the local particle's mean collision time.
The UGKWP method is also extended to the
coupled radiation-material system. With the inclusion of energy
exchange, the UGKWP method can give excellent simulation results in
different regimes.  A few benchmark problems are tested to show the
performance of the current scheme.  The accuracy and efficiency of the
UGKWP method are fully confirmed.
For the multiscale transport, in the cases with the co-existence of different regimes the UGKWP may improve the efficiency
on several-order-of-magnitude in comparison with the purely particle methods.
The UGKWP takes also advantages of both particle and macroscopic solver. The UGKWP has totally removed the ray effect and
and has a much improved efficiency in comparison with DOM-type UGKS.
Based on the direct modeling of the transport physics in the time step scale \cite{xu2015direct}, such as the relationship between $t_c$ and 
$\Delta t$ in the current study, we have three versions of UGKS for the multiscale transport simulations, i.e., the DOM or DVM type,
the purely particle formulation, and the wave-particle one.
The advantages and disadvantages of these different discretization under the same UGKS framework will be further investigated,
and the extension of these schemes to complex systems, such as radiative-hydrodynamics, plasma, multi-component system, and reactive flow,
will be constructed.

\section*{Acknowledgement}

The current research is supported by Hong Kong research grant council
(16206617),  National Science Foundation of China (11472219,11772281,91852114,11771035, U1530401).


\bibliography{references}

\end{document}